\documentstyle[11pt,amssymb,amsfonts,amsthm,amssymb,amscd,amstext,epsfig]{article}

\def\ben{\begin{equation}}
\def\beq{\begin{equation}}
\def\een{\end{equation}}
\def\eeq{\end{equation}}
\let\a=\alpha \let\b=\beta  \let\d=\delta \let\e=\epsilon
  \let\q=\theta 
    \let\p=\phi 
\let\s=\sigma   

\let\w=\omega

\let\la=\label  
  
\def\nn{\nonumber}

\let\pa=\partial
\let\6=\partial
\def\be{\begin{equation}}
\def\ee{\end{equation}}
\def\ba{\begin{array}}
\def\ea{\end{array}}
\def\del{\partial}

\def\dalemb#1#2{{\vbox{\hrule height .#2pt
        \hbox{\vrule width.#2pt height#1pt \kern#1pt
                \vrule width.#2pt}
        \hrule height.#2pt}}}

\newcommand{\bea}{\begin{eqnarray}}
\newcommand{\eea}{\end{eqnarray}}
\newcommand{\tr}{{\rm tr} }

\def\R{{{\Bbb R}}}

\def\Z{{{\Bbb Z}}}

\def\ih{{\bar{\imath}}}

\def\nh{{\bar{n}}}
\def\mh{{\bar{m}}}
\def\uh{{\bar{u}}}
\def\vh{{\bar{v}}}
\def\sh{{\bar{s}}}
\def\th{{\bar{t}}}
\def\zh{{\bar{\psi}}}
\def\rh{{\bar{r}}}

\def\Ncal{{\mathcal{N}}}
\def\bmet{{{\mathbb B}}_7}
\def\cmet{{{\mathbb C}}_7}
\def\ctmet{\widetilde{{{\mathbb C}}_7}}
\def\dmet{{{\mathbb D}}_7}

\def\st{\sin\theta}
\def\ct{\cos\theta}
\def\sa{\sin\alpha}
\def\ca{\cos\alpha}
\def\raisenot{\raise .5mm\hbox{/}}
\def\slashD{\hbox{{$D$}\kern-.61em\hbox{\raisenot}}}
\def\O{{{\mathcal{O}}}}
\newcommand{\eps}{\epsilon}

\begin{document}
%%%%%%%%%%%%%%%% title page %%%%%%%%%%%%%%%%%%%%%%%%%%%%%%%%%%%%
\begin{titlepage}
\vfill
\begin{flushright}
MIT-CTP-3435\\
DAMTP-2003-122\\
hep-th/0311088\\
\end{flushright}
\vskip 1cm
\begin{center}
\baselineskip=16pt
{\LARGE\bf The chiral anomaly from M theory}
\vskip 0.3cm
{\Large {\sl }}
\vskip 1cm
{\bf ~Umut G\"ursoy${}^{\star}${}\footnote{e-mail: umut@mit.edu},
     ~Sean A. Hartnoll${}^{\dagger}${}\footnote{e-mail: s.a.hartnoll@damtp.cam.ac.uk}
 and ~Rub\'en Portugues${}^{\dagger}${}\footnote{e-mail: r.portugues@damtp.cam.ac.uk}}\\
\vskip 1cm
{\small
${}^{\star}$
Center for Theoretical Physics\\
Massachusetts Institute of Technology\\
Cambridge MA 02139, USA\\
\vskip 0.4cm
${}^{\dagger}$
DAMTP, University of Cambridge, \\
Centre for Mathematical Sciences,
Wilberforce Road, \\
Cambridge CB3 0WA, UK\\
}
\end{center}
\vskip 1cm

\begin{abstract}
We argue that the chiral anomaly of $\Ncal = 1$ super Yang-Mills
theory admits a dual description as spontaneous symmetry breaking
in M theory on $G_2$ holonomy manifolds. We identify an angle of
the $G_2$ background dual to the anomalous $U(1)_R$ current in
field theory. This angle is not an isometry of the metric and we
therefore develop a theory of ``massive isometry'' to describe
fluctuations about such angles. Another example of a massive
isometry occurs in the Atiyah-Hitchin metric.
\end{abstract}

\end{titlepage}
%%%%%%%%%%%%%%%%%%%%%%%%%%%%%%%%%%%%%%%%

\setcounter{equation}{0}
\setcounter{page}{1}
\enlargethispage{2mm}
\tableofcontents
\addtocontents{toc}{\protect\setcounter{tocdepth}{3}}
\vfill\eject

\section{Background and motivation}

Dualities between nonconformal ${\mathcal{N}}=1$ gauge theories
and gravity theories on supersymmetric backgrounds continue to
provide important insight into the dynamics of strongly coupled
gauge theories with similar properties to Quantum Chromodynamics.

In this paper we explore these dualities further. Specifically, we
work towards placing dualities involving $G_2$ holonomy M theory
backgrounds on a comparable footing to those introduced by
Klebanov-Strassler \cite{Klebanov:2000hb} and
Maldacena-N\'u\~{n}ez \cite{Maldacena:2000yy}. An important
property of $G_2$ holonomy solutions is that classically they are
purely gravitational, with no fluxes.

We will match a dynamical phenomenon on the gravity and gauge
theory sides of the duality. Namely, we argue that the chiral
anomaly of ${\mathcal{N}}=1$ super Yang-Mills (SYM) theory is dual
to spontaneous symmetry breaking in the $G_2$ background. In order
to make this matching we introduce the concept of a ``massive
isometry'' of a gravity background. We develop this idea further
in the later parts of the paper.

Many questions require further investigation, we mention
some of these in the final discussion.

\subsection{${\mathcal{N}}=1$ backgrounds and the $G_2$ holonomy challenge}

The original Anti-de Sitter / Conformal Field Theory (AdS/CFT)
correspondence \cite{Maldacena:1997re,Gubser:1998bc,Witten:1998qj}
showed that the idea of a duality between gauge theory and gravity
could be realised by critical string theory. Given this, an
important question is whether string theory can provide dual
descriptions to strongly coupled nonconformal theories with
reduced supersymmetry.

Initial progress in this direction came through understanding
deformations of ${\mathcal{N}}=4$ field theory
\cite{Girardello:1998pd,Distler:1998gb}, see \cite{Petrini:1999qa}
for a review. Deformations involve either nonzero vacuum
expectation values in the field theory or the addition of relevant
or marginal operators to the action. The resulting five
dimensional gravity backgrounds, after dimensional reduction on
any compact internal space, are asymptotic to AdS spacetime,
corresponding to the deformations becoming negligible in the
ultraviolet (UV) of the field theory.

These cases were then generalised by the `geometrical engineering'
within string theory of gravity duals to field theories without a
UV conformal fixed point. Two completely regular backgrounds are
known, which provide duals to nonconformal ${\mathcal{N}}=1$ field
theories. One is the Klebanov-Strassler background
\cite{Klebanov:2000hb}, constructed using fractional branes placed
at a conifold singularity. The other is the Maldacena-N\'u\~{n}ez
background \cite{Maldacena:2000yy}, constructed as a
supersymmetric configuration of D5-branes wrapped on an $S^2$
inside a Calabi-Yau manifold. A recent clear review of
nonconformal dualities with ${\mathcal{N}}=1$ and
${\mathcal{N}}=2$ supersymmetry is \cite{Bigazzi:2003ui}. From a
ten dimensional string theory perspective, the duality we
investigate here is conceptually similar to a IIA version of the
Maldacena-N\'u\~{n}ez background, where now we have D6-branes
wrapping an $S^3$. However, there does not appear to be a direct
connection via string theory dualities between the backgrounds we
consider and the Maldacena-N\'u\~{n}ez background.

It is important to check the genericity of predictions from
gravity duals by studying new examples with different properties.
As we just mentioned, an alternative way to geometrically engineer
${\mathcal{N}}=1$ SYM theory is to wrap D6-branes on an $S^3$
inside a Calabi-Yau manifold \cite{Vafa:2000wi}. An advantage of
this setup, compared with the IIB setup of wrapped D5-branes, is
that the resulting configurations have a very simple lift to M
theory: they are described by purely gravitational backgrounds
with $G_2$ holonomy \cite{Atiyah:2000zz}. The M theory perspective
proved useful in clarifying the string theory duality of
\cite{Vafa:2000wi} in terms of a topology change in M theory. This
process was further systematised at the quantum level by Atiyah
and Witten \cite{Atiyah:2001qf} who were able to control quantum
corrections to show that the topology change was smooth. The
connection between ${\mathcal{N}}=1$ SYM theory and $G_2$ holonomy
manifolds was further developed by studying the structure of ADE
singularities in such manifolds
\cite{Acharya:1998pm,Acharya:2000gb}.

In parallel with the development of physical intuition associated
with $G_2$ holonomy, there has been important progress recently in
constructing new explicit cohomogeneity one $G_2$ metrics. The
original construction of $G_2$ metrics in \cite{bs,Gibbons:er} was
generalised to cases where the dimensionally reduced IIA dilaton
remained finite asymptotically \cite{Brandhuber:2001yi}. This was
later further generalised to cases which shared the improved
asymptotic behaviour and further had no orbifold singularity at the
origin \cite{Brandhuber:2001kq,Cvetic:2001kp,Cvetic:2001ih},
corresponding perhaps to backgrounds giving a good infra-red (IR)
description of gauge theory physics.

Despite such major progress regarding $G_2$ holonomy, the
understanding of a relationship with ${\mathcal{N}}=1$ gauge
theory and the construction of explicit metrics have not yet
coalesced into the development of a duality along the lines of the
Klebanov-Strassler or Maldacena-N\'u\~{n}ez backgrounds.

Various topological quantities have been matched between the
gravity and field theory descriptions. Thus wrapped membranes in
the IR geometry describe gauge theory confining strings
\cite{Acharya:2000gb,Acharya:2001hq} and wrapped fivebranes on
three-cycles describe gauge theory domain walls
\cite{Acharya:2000gb,Acharya:2001dz}. However, the $G_2$ metrics
themselves, which will be needed to match dynamical phenomena, are
not used in these calculations. Some properties of $G_2$ metrics
were used by Atiyah and Witten \cite{Atiyah:2001qf} in calculating
the space of asymptotically $G_2$ vacua, although the metrics used
were old asymptotically conical cases \cite{bs,Gibbons:er} that do
not admit a IIA reduction with everywhere finite dilaton. In
contrast, the physics we describe will depend crucially on the
leading and subleading asymptotic form of the recently constructed
metrics.

As far as dynamics are concerned, a recent study of rotating
membranes in explicit $G_2$ backgrounds \cite{Hartnoll:2002th}
reproduced a logarithmic relationship between spin and energy that
is characteristic of twist two operators in gauge theories.
However, it was not clear how to concretely match this tantalising
result with gauge theory quantities.

We argue here that the chiral anomaly in field theory admits a
dual description as spontaneous symmetry breaking of a gauge
symmetry in M theory. This enables us to explicitly identify a
circle in the recently constructed $G_2$ metrics that is dual to
the chiral symmetry of the field theory.

\subsection{The chiral anomaly in gauge theory}

Let us briefly recall the origin of the chiral anomaly in
${\mathcal{N}} = 1$ Yang-Mills theory. We follow the succinct and
fully nonperturbative exposition of \cite{seiberg}.

Pure ${\mathcal{N}} = 1$ Yang-Mills theory in four dimensions
contains a massless spin-half gluino field, $\Psi$, that
transforms in the adjoint representation of the gauge group. The
Lagrangian
\be
\frac{i}{g^2} {\bar{\Psi}} \slashD_{\text{adj}} \Psi ,
\ee
is classically invariant under the $U(1)_R$ chiral symmetry
\be\label{eq:Rsym}
\Psi \to e^{i \a \gamma_5} \Psi,
\ee
where as usual $\gamma_5$ has eigenvalues $\pm 1$ acting on
positive and negative chirality spinors respectively, and $0 \leq
\a < 2\pi$ parametrises the $U(1)$ group. The notation $U(1)_R$
means that the chiral symmetry in this case is the R symmetry of the
${\mathcal{N}} = 1$ supersymmetry algebra.

The quantum mechanical anomaly may be seen from the nontrivial change
in the path integral measure under the transformation (\ref{eq:Rsym})
\be
D \Psi \to \frac{D \Psi}{\det e^{i \a \gamma_5}} = D \Psi
e^{-i\a\tr\gamma_5} = D \Psi e^{-i \a \text{Ind}
\slashD_{\text{adj}}} .
\ee
The steps used here are as follows. The first step is simply the
Jacobian of the chiral transformation
(\ref{eq:Rsym}). The second uses the standard relation between the
determinant and trace of operators. Note that the trace of
$\gamma_5$ is not over spinor matrix indices but rather over state
space. The final step uses the fact that the Dirac index is
defined as the difference between the number of positive and
negative chirality massless modes, and that massive modes always
come in pairs of positive and negative chirality and so do not
contribute to the trace of $\gamma_5$.

If the gauge group is $SU(N)$, the index of the Dirac operator in the adjoint
representation is an integral multiple of $2N$.
Therefore, the surviving symmetry is $\Z_{2N}$, corresponding to
values $\a = (2 \pi n)/(2N)$, with $n \in 0 \ldots 2N$.

We have included this short derivation to emphasise the fact that
a nonperturbative treatment is possible. One could obtain the same
results from the standard perturbative one-loop triangle diagrams.
The protection of the chiral anomaly from nonperturbative
corrections is ultimately why the phenomenon is amenable to study
in dualities.

\subsection{Dual descriptions of the chiral anomaly}

A dual description of the anomaly in the chiral $SU(4)_R$ symmetry
in ${\mathcal{N}} = 4$ super Yang-Mills, with external gauged $SU(4)$ sources,
was amongst the first
achievements of the AdS/CFT duality \cite{Witten:1998qj}. The
principal steps in the matching are as follows. Firstly, that
the gravity background has a gauge field excitation $A_{\mu}$ that
couples to the chiral symmetry current $J^{\mu}$ of the dual theory. Secondly,
the anomaly in the field theory, $<\pa \cdot J> \neq 0$, translates into
a breaking of gauge invariance in the gravity solution.

In the ${\mathcal{N}} = 4$ case, the breaking of gauge invariance
in the gravity solution occurs due to boundary terms that arise
because of a Chern-Simons term in the supergravity action
\cite{Witten:1998qj}. The only other known mechanism of breaking
gauge invariance in a physically sensible way is through
spontaneous symmetry breaking.

Spontaneous symmetry breaking was first related to the chiral symmetry in the
study of deformations of the AdS/CFT correspondence
\cite{Bianchi:2000sm,Brandhuber:2000fr}. The gravity duals are described
by a five dimensional gauged supergravity in an asymptotically AdS
background. In the case of flow along the Coloumb branch of the
${\mathcal{N}} = 4$ theory, one sees that the R-symmetry $SO(6) =
SU(4)$ is broken in the interior of the spacetime to $SO(p)$, $p\leq6$, by
some of the $SO(6)$ gauge fields acquiring a mass
\cite{Brandhuber:2000fr}. In these cases, the breaking of chiral symmetry in
the field theory is not due to anomaly. Instead, the symmetry is explicitly broken
by vacuum expectation values or by deformations.

The use of spontaneous symmetry breaking as dual to breaking of
chiral symmetry was extended to anomalous, as opposed to explicit,
breaking in \cite{Klebanov:2002gr,Krasnitz:2002ct}. The duality in
question involved the Klebanov-Strassler ${\mathcal{N}}=1$
solution. The key idea, reviewed in the next subsection, works as
follows. The asymptotic, UV, metric has a $U(1)$ isometry. When
this direction is perturbed, the Lagrangian for the fluctuation is
described by the usual gauge-kinetic term, $F^2$, for a vector
field. However, in order for the perturbation to be consistent,
the background Ramond-Ramond fields must also be perturbed, as
well as the metric. These contribute an extra mass term to the
Lagrangian for the fluctuation. The resulting Lagrangian for a
massive vector field may be understood as the result of
spontaneous symmetry breaking. In the spontaneous breaking, the
gauge field `eats' a scalar field to become massive. The scalar
field itself does not acquire a nonzero vacuum expectation value
in order to break the symmetry. Strictly speaking the scalar is
thus not a Higgs field, but rather a St\"{u}ckelberg scalar field.
Reproducing this phenomenon for the $G_2$ holonomy duality, where
there are no p-form field strengths, will be the main objective of
this work.

Another aspect of the chiral anomaly that is visible in the
Klebanov-Strassler and Maldacena-N\'u\~{n}ez backgrounds is the
breaking of $U(1)$ to $\Z_{2N}$ due to field theory anomaly in the
UV, and then the breaking of $\Z_{2N}$ to $\Z_2$ in the IR
\cite{Maldacena:2000yy,Klebanov:2002gr,DiVecchia:2002ks}.

\subsection{Spontaneous symmetry breaking in the Maldacena-N\'u\~{n}ez background}
\label{sec:SSBMN}

Rather than review literally the argument of
\cite{Klebanov:2002gr,Krasnitz:2002ct} for the Klebanov-Strassler
background, we show that it is trivially adapted to the
Maldacena-N\'u\~{n}ez (MN) background \cite{Maldacena:2000yy}. In
fact it works more cleanly, because there is no self-dual five
form. The result is interesting in its own right because it
provides an understanding of the chiral symmetry breaking in the
MN solution as spontaneous symmetry breaking due to non-invariance
of the Ramond-Ramond (RR) two-form potential under the chiral
isometry.

The MN solution considers D5-branes wrapping the topologically non
trivial two-cycle, $S^2$, within the resolved conifold, a Calabi-Yau
three-fold. In the spirit of the gravity/gauge theory correspondence,
the field theory on the D-brane worldvolume in a certain regime,
is dual to the supergravity solution generated by the gravitating
D-branes. For the MN configuration, the field theory is ${\mathcal
N}=1$ SYM theory. When the backreaction of the branes is taken
into account, the two-cycle, $S^2$, becomes topologically trivial
and a three-cycle, $S^3$, blows up. Thus the dual geometry is the
deformed conifold with non-vanishing fluxes.

Since the anomalous breaking of chiral symmetry is a UV effect in
field theory, consider the geometry at large $r$, in the Einstein
frame and with $\alpha^{\prime} = 1$ :
\bea
ds_{10}^2 &=& e^{\Phi/2}\bigg[dx_{1,3}^2 + N \bigg(dr^2 + r
[d\theta^2
+ \sin^2\theta d\varphi^2] + \nn \\
& & +\frac{1}{4} [d\alpha^2 +
\sin^2\alpha d\beta^2 + (d\psi + \cos\alpha d\beta - \cos\theta
d\varphi)^2]\bigg)\bigg]\label{MNUV} \,,
\eea
where the dilaton at large $r$ is,
\be
e^{2\Phi}=e^{2\Phi_0}\frac{e^{2r}}{4\sqrt{r}}.
\label{dil}
\ee
There is the following flux
\be
\label{3form}
F_3  =  -\frac{1}{4} N \biggl[ \omega_2 \wedge d\psi -
  \ct \sa \, d\alpha\wedge d\beta\wedge d\varphi + \st \ca \,
  d\theta\wedge d\varphi \wedge d\beta \biggl]\, ,
\ee
where $\omega_2$ is the volume form on a two-cycle
\be
\omega_2 = \sin\alpha \, d\alpha\wedge d\beta  - \st \, d\theta\wedge
  d\varphi.
\ee
The RR two-form potential is
\be\label{RRpot}
C_2 = \frac{1}{4} N \biggl[ \psi\,\, \omega_2  +
  \cos\theta \, \cos\alpha\, d\varphi\wedge d\beta  \biggl].
\ee

The metric (\ref{MNUV}) clearly exhibits the isometry
\be\label{iso}
\psi\to\psi+\eps.
\ee
In the same spirit as \cite{Klebanov:2002gr} we propose this
isometry as the dual of the R-symmetry in the UV limit of
${\mathcal{N}}=1$ SYM and call it the ``chiral isometry'' from now
on. This is a symmetry of the RR field strength (\ref{3form}),
although not of the potential (\ref{RRpot}). In section
\ref{sec:z2n} below, this fact will result in the breaking
$U(1)\to \Z_{2N}$.

We will consider a fluctuation about the chiral isometry. The
presence of the field strength will result in the usual vector
perturbation obtaining a mass through spontaneous breaking. This
is crucial for the gauge theory/gravity correspondence because
breaking of a global symmetry in the field theory by any means
(explicit, anomalous or spontaneous) should be sought as a
spontaneous symmetry breaking of the dual gauge symmetry on the
gravity side.

The perturbation of the metric (\ref{MNUV}) and potential
(\ref{RRpot}) is
\be
d\psi \to d\psi + A_a(x,r) dx^a \,, \qquad \psi \to \psi +
\lambda(x,r) ,
\ee
where the index $a$ runs over the five coordinates $t,x,y,z,r$.
Introducing this perturbation is just a gauging of the chiral
isometry, because now the coordinate transformation $\psi \to \psi
+ \e(x,r)$ corresponds to the $U(1)$ gauge transformations
\be\label{eq:gauge1}
A(x,r) \to A(x,r) + d\e(x,r) \,, \qquad \lambda(x,r) \to
\lambda(x,r) + \e(x,r) .
\ee
Note that we needed to perturb the RR potential in order to gauge
the isometry.

As usual, the Ricci scalar of the perturbed metric is given the
original value plus a gauge-kinetic term
\be\label{ricciA}
R(A)=R(A=0) -\frac{1}{16} N e^{\Phi/2} |F_2|^2\,,
\ee
where $F_2 = dA$. The three-form field strength becomes
\be
F_3(\lambda) = F_3(\lambda=0) - \frac{1}{4} N \, (\omega_2
\wedge d\lambda)\,.
\ee

At this point, one should introduce the St\"{u}ckelberg-Pauli
vector field $W=A-d\lambda$ which is gauge invariant. In terms of
the gauge invariant quantity $W$ we obtain
\be
|F_3(\lambda)|^2 = |F_3(0)|^2 + \frac{3(1+16r^2)}{8 r^2 N}
e^{-\Phi} N \, W_a W^a \,.
\label{tildeF3}
\ee
Note that the calculation is rather remarkable.  All the
components of the gauge field $A$ come from the metric, while the
components of $\lambda$ come from the three-form. They rearrange
themselves exactly to produce the mass term for $W$. The gauge field
has `eaten' the scalar fluctuation to become massive.

The type IIB action in the Einstein frame for the relevant fields,
which yields the MN solution, is
\be
S_{\text{IIB}} \propto \int d^{10}x \sqrt{g_{10}} \left[R -
\frac{1}{2} e^{\Phi} |F_3|^2 \right] .
\label{actionstring}
\ee
Plugging (\ref{ricciA}) and (\ref{tildeF3}) into this action one
obtains the action for the perturbation
\be
S\propto - \int d^{10}x \sqrt{g_{10}}
\left[\frac{N}{16} e^{\Phi/2} |F_2|^2 + \frac{3(16r^2 +1)}{16r^2}W^2\right],
\label{actionE}
\ee
where we have only shown the relevant terms for the discussion,
dropping an overall constant term. Here $F_2 = d W$. One sees that
a mass term for the gauge field $W$ arises as a result of the
spontaneous breaking of the gauge symmetry. Instead of $W^2$ in
the action (\ref{actionE}), one could have written $(D\lambda)^2$,
with the appropriate covariant derivative $D\lambda = \pa\lambda -
A$. Although the scalar field $\lambda$ need not acquire a nonzero
vacuum expectation value, its presence still implies spontaneous
symmetry breakdown. This is because even the vacuum $<\lambda> =
0$ is not invariant under the transformation $\lambda
\to \lambda + \eps$.

There is no problem here with substituting the perturbation ansatz
into the action. The gauge transformations (\ref{eq:gauge1}) allow
us to consider the perturbation in a gauge where $\lambda = 0$.
Such perturbations about an isometry are known to be consistent,
that is, the equations of motion obtained from the action
(\ref{actionE}) are the correct equations of motion for the
perturbation. In the $G_2$ holonomy cases we consider below,
checking consistency will be an important and nontrivial test of
the scenario we present for spontaneous symmetry breaking.

\section{Spontaneous symmetry breaking in $G_2$ backgrounds}

{\it In this section, we show the spontaneous symmetry breaking of
a $U(1)$ fluctuation about $G_2$ holonomy backgrounds.}

\subsection{The $G_2$ holonomy metrics}

All known $G_2$ holonomy metrics that are cohomogeneity one with
principal orbits given by $SU(2)\times SU(2) = S^3\times S^3$ admit the following
form \cite{Cvetic:2001kp,Chong:2002yh}. At the end of this subsection
we comment on the topology of these spaces and give some illustrations.

\bea\label{eq:sigmametric}
ds^2_{11} = dx_4^2 + dr^2 + a(r)^2\left[(\Sigma_1 + g(r)\s_1)^2
+(\Sigma_2 + g(r)\s_2)^2\right]+ f(r)^2(\Sigma_3 + g_3(r) \s_3)^2  \nn \\
+ b(r)^2\left[(\Sigma_1 - g(r)\s_1)^2
+(\Sigma_2 - g(r)\s_2)^2\right]  + c(r)^2(\Sigma_3 - \s_3)^2,
\eea
where $\Sigma_i,\s_i$ are left-invariant one-forms on two copies of $SU(2)$,
\bea\label{eq:1forms}
\s_1 & = & \cos\psi_1 d\q+\sin\psi_1\sin\q d\phi , \nn \\
\s_2 & = & -\sin\psi_1 d\q+\cos\psi_1\sin\q d\phi , \nn \\
\s_3 & = & d\psi_1 + \cos\q d\phi ,
\eea
where $0\leq\q\leq\pi$, $0\leq\phi\leq 2\pi$, $0\leq\psi_1\leq
4\pi$, at least before including the ${\mathbb{Z}}_N$ quotient,
which we describe below. The definitions for $\Sigma_i$ are
analogous but with $(\q,\phi,\psi_1) \to (\alpha,\beta,\psi_2)$.
Note that here and throughout we work in units with the eleven
dimensional Planck length $l_p = 1$.

The radial functions are defined through a constraint
\be
g_3 =  g^2 - \frac{ c\, ( a^2- b^2) (1- g^2)}{2 a\,  b\,  f}\,,
\ee
and the following first order equations
\bea\label{eq:firstorder}
\dot{a} & = & \frac{ c^2\, ( a^2 - b^2) +
[4 a^2\, ( a^2- b^2)-  c^2\, (5  a^2- b^2) - 4
a\,  b\,  c\,  f]\,  g^2}{16 a^2\,  b\,  c\,
g^2}\,,\nn\\
\dot{b} & = & -\, \frac{ c^2\, ( a^2- b^2) + [4  b^2\,
( a^2 - b^2) + c^2\, (5 b^2 -  a^2) - 4 a\,  b\,
 c\,  f]\, g^2}{16  a\,  b^2\,  c\,  g^2}\,,\nn\\
\dot{c} & = & \frac{ c^2 + ( c^2 -2 a^2 -2 b^2)\,
g^2}{4 a\,  b\,  g^2}\,, \nn \\
\dot{f} & = & -\, \frac{( a^2- b^2)\, [ 4  a\,  b\,  f^2\,
 g^2 -  c\, (4 a\,  b\,  c +  a^2\,  f -
b^2\,  f)\, (1- g^2)]}{16  a^3\,  b^3\,  g^2}\,,\nn\\
\dot{g} & = & -\, \frac{ c\, (1- g^2)}{4 a\,  b\,  g} \,.
\eea
These equations have two known integration constants
\be\label{eq:conserved}
m = (b^2 - a^2) c g^2 + 2 a b f g^2 g_3 \,, \qquad n = (b^2-a^2)c +
2 a b f .
\ee

It is more convenient here to rewrite these metrics in a form that
makes manifest the isometry along the coordinate $\gamma = \psi_1 +
\psi_2$ and which completes
the square in the tentative `massive isometry' coordinate $\psi =
\psi_1 - \psi_2$.
\be\label{eq:g2metric}
ds^2_{11} = dx_4^2 + dr^2 + a^2 \left[(g^1)^2 + (g^2)^2 \right] +
b^2 \left[(g^3)^2 + (g^4)^2 \right] + u^2 (g^5)^2 + v^2 (g^6)^2 ,
\ee
where we have introduced the new radial functions
\be
u^2 = \frac{f^2 (1-g_3)^2}{4} + c^2  \,, \qquad v^2 =
\frac{c^2 f^2 (1+g_3)^2}{4 u^2}\,,
\ee
and also the terms
\bea
g^1 & = & g(r) E^1 + E^3 \equiv - g(r) \sin\theta d\phi  - \sin\psi d\a
- \cos\psi\sin\a d\b \,, \nonumber \\
g^2 & = & g(r) E^2 + E^4 \equiv g(r) d\theta + \cos\psi d\a
- \sin\psi\sin\alpha d\b \,, \nonumber \\
g^3 & = & g(r) E^1 - E^3 \,, \nonumber \\
g^4 & = & g(r) E^2 - E^4 \,, \nonumber \\
g^5 & = & d\psi + p(r) d\gamma + q(r)
\cos\theta d\phi + s(r) \cos\alpha d\beta \,,
\nonumber \\
g^6 & = & d\gamma + \cos\theta d\phi + \cos\alpha d\beta \,.
\eea
Note that, although similar, $E^3$ and $E^4$ are not the standard conifold
terms that were used to write a $G_2$ holonomy metric in
\cite{Brandhuber:2001yi}. The standard case would be recovered if we were to
work with $\b^{\prime} = -\b$. The further new radial functions introduced are
\be
p = \frac{f^2 (g_3^2-1)}{4 u^2}\,, \qquad  q = \frac{2 f^2 g_3 (g_3-1)+4c^2}{4 u^2}\,, \qquad
s = \frac{2 f^2 (g_3-1)-4c^2}{4 u^2}\,.
\ee
The angles have ranges $0 \leq \a \,, \theta \leq \pi$, $0 \leq \b
\,, \phi < 2\pi$, $0 \leq \psi < 4\pi $ and $0 \leq \gamma < 4\pi/N$.
These variables make the $\pa_\gamma$ isometry explicit. In the
range of $\gamma$ we have now included the quotient by $\Z_N$: $\gamma
\sim \gamma + 4\pi/N$. The direction $\pa_\gamma$ is usually thought
of as the M theory circle. The quotient is important because on
reduction to IIA, it will give $N$ units of D6-brane flux.
Classically, it is the only place in the M theory background where
the $N$ of the dual $SU(N)$ gauge theory appears.

General solutions to these equations are only known numerically.
Two nonsingular exact solutions are known
\cite{bs,Gibbons:er,Brandhuber:2001yi,Cvetic:2001bw}, but will not
be of particular interest here. We will work directly with the
equations (\ref{eq:firstorder}). These first order equations imply
$G_2$ holonomy of the metric. They further imply the set of second
order equations for Ricci flatness of the metric
(\ref{eq:g2metric}). The second order equations allow more
general, non-supersymmetric solutions. However, we will see that
in considerations of consistency below, it is the first order
equations that are required.

The equations (\ref{eq:firstorder}) may be specialised to various
cases that have been discussed in the literature. The unified form
we use here allowed a classification of the known $G_2$ holonomy
metrics with $SU(2)\times SU(2)$ principal orbits
\cite{Cvetic:2001kp,Cvetic:2001ih}. Four families were discussed,
denoted $\bmet$ \cite{Brandhuber:2001yi,Cvetic:2001zx} , $\cmet$
\cite{Cvetic:2001sr}, $\ctmet$ \cite{Cvetic:2001kp} and $\dmet$
\cite{Brandhuber:2001kq,Cvetic:2001ih}. The first fact that we
shall use about these different families is that $\bmet$ and
$\cmet$ metrics follow from making use of the consistent
truncation of the first order equations $g(r)=1$. However,
we will ultimately argue that the $\dmet$ metrics are the
gravitational dual of ${\mathcal{N}}=1$ SYM field theory.

The $\bmet$ and $\dmet$ families will appear to be the most closely
related to ${\mathcal{N}}=1$ SYM. The $\bmet$ metrics have topology
$S^3\times (\R^4/\Z_N)$, and are singular, whilst the $\dmet$ metrics
have nonsingular topology $S^3/\Z_N \times \R^4$. Both families have
the same asymptotics. The topology is illustrated in the following
figure.

\begin{figure}[h]
\begin{center}
\vspace{4mm}
\epsfig{file=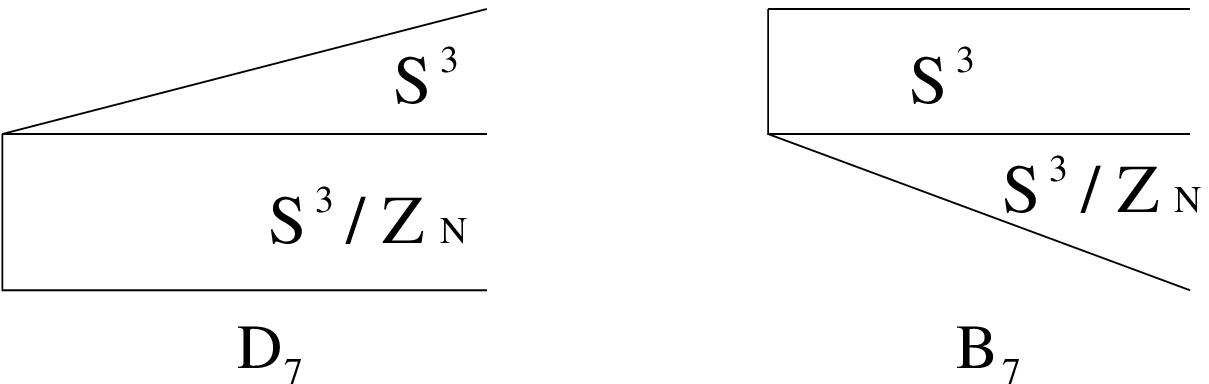,width=8cm}

\noindent {\bf Figure 1:} The $\bmet$ and $\dmet$ spaces.
\end{center}
\end{figure}

The geometric differences between the distinct families
may also be seen through a dimensionally reduced
description in terms of four dimensional nonabelian monopoles and
cosmic strings \cite{Hartnoll:2001qa}.

\subsection{A fluctuation about the $G_2$ metrics}
\label{sec:g2fluctuation}

We will argue that the direction $\pa_{\psi}$ is dual to the
anomalous chiral symmetry of ${\mathcal{N}} = 1$ SYM gauge theory.
This angle is not a $U(1)$ isometry of the metric. It does
contain, however, a $\Z_2$ isometry of the metric, under $\psi \to
\psi + 2\pi$. Therefore it is a natural candidate for the dual to
the chiral symmetry, because chiral symmetry breaking should
preserve a $\Z_2$ symmetry all the way into the IR. Another
argument is that whilst not an isometry, this direction is still
special in a way that will be formalised in a later section where
we will call such a direction a massive isometry. Further,
$\pa_{\psi}$ becomes an isometry at infinity.

Fluctuations about an isometry are well known to be described by
gauge fields. Fluctuations about a massive isometry are described
by gauge fields that obtain a mass through spontaneous symmetry
breaking.

As described in the introduction, spontaneous symmetry breaking in
supergravity solutions is the most natural way to break gauge
symmetries that are dual to broken global symmetries in field
theory. In particular, spontaneous symmetry breaking is known to
provide the dual description to the chiral anomaly in the
Klebanov-Strassler \cite{Klebanov:2002gr,Krasnitz:2002ct} and
Maldacena-N\'u\~{n}ez, see introduction, solutions. A novel
feature of the present case is that the background is purely
gravitational.

Consider the following fluctuation about the metric
(\ref{eq:g2metric}) in the $\pa_{\psi}$ direction
\be\label{eq:fluctuation}
d\psi \to d\psi + A_a(x,r) dx^a \,, \qquad \psi \to \psi +
\lambda(x,r) ,
\ee
where the index $a$ runs over the five coordinates $t,x,y,z,r$.

The coordinate transformation $\psi \to \psi + \e(x,r)$
corresponds to the $U(1)$ gauge transformations
\be\label{eq:gauge}
A \to A + d\e \,, \qquad \lambda \to \lambda + \e .
\ee

Substituting this fluctuation into the eleven dimensional
Einstein-Hilbert action, one obtains the following reduced action
for the fluctuation
\be\label{eq:Aaction}
S \propto - \int d^4x\,dr \sqrt{g_7(r)}
\left[ \left(\frac{(g^2-1)^2 [2abf + c(a^2-b^2)]^2}{64 a^2 b^2} + \frac{c^2}{4} \right) F_{a b}
  F^{a b} + \frac{[a^2-b^2]^2}{4a^2 b^2} D_a \lambda D^a \lambda \right] ,
\ee
where the field strength is as usual
\be
F_{ab} = \pa_a A_b - \pa_b A_a ,
\ee
the covariant derivative is that corresponding to the
representation of $\lambda$
\be
D_a \lambda = \pa_a \lambda - A_a ,
\ee
and $g_7(r)$ is the determinant of the $G_2$ metric with the
angular directions integrated out,
\be\label{eq:determinant}
g_7(r) = a^2 b^2 c^2 g^4 \left[2abf(1+g^2) + c(b^2-a^2)(1-g^2)
\right]^2 .
\ee
Indices are raised using the flat five-dimensional part of the
metric (\ref{eq:g2metric}), that is
\be
ds^2_5 = dx^2_4 + dr^2 .
\ee

It is clear that the action is invariant under the gauge
transformations (\ref{eq:gauge}). The derivation of the action
(\ref{eq:Aaction}) only uses the equations for the radial
functions (\ref{eq:firstorder}) in order to set a cosmological
term to zero, corresponding to Ricci-flatness of the eleven
dimensional metric.

The reduced action is naturally interpreted as exhibiting
spontaneous symmetry breaking, as we discussed towards the end of
section \ref{sec:SSBMN}. The scalar field $\lambda$ need not
acquire a nonzero vacuum expectation value. The gauge field $A$
`eats' the scalar field $\lambda$ to become a gauge-invariant
massive St\"{u}kelberg-Pauli vector field
\be
W = A - d\lambda .
\ee
In terms of this field, the action becomes
\be\label{eq:Waction}
S \propto - \int d^4x\,dr \sqrt{g_7(r)}
\left[ \left(\frac{(g^2-1)^2 [2abf + c(a^2-b^2)]^2}{64 a^2 b^2} + \frac{c^2}{4} \right) F_{a b}
  F^{a b} + \frac{[a^2-b^2]^2}{4a^2 b^2} W_a W^a \right] ,
\ee
where now
\be
F_{ab} = \pa_a W_b - \pa_b W_a \, \left( = \pa_a A_b - \pa_b A_a
\right) .
\ee
Thus vector fluctuations about the angle $\pa_{\psi}$ do acquire a
mass through spontaneous symmetry breaking, as required for the
dual description of the anomalous chiral symmetry.

Let us note in passing that the action is greatly simplified in
the case when $g(r)=1$, corresponding to the $\bmet$ and $\cmet$
metrics. In this case
\be\label{eq:gisoneaction}
S_{g=1} \propto - \int d^4x\,dr
\left[ a^2 b^2 c^3 f  F_{a b}
  F^{a b} + (a^2-b^2)^2 c f W_a W^a \right] .
\ee

It is important to check that this reduction is consistent, that
is, that the corresponding reduced equations of motion solve the
initial eleven dimensional equations of motion. We do this in the
following section. Beforehand, we make a few additional comments about
why it was necessary to consider a non-isometric angle.

The most immediate reason, which we have already stated, is that we
wish to exhibit spontaneous symmetry breaking. In the absence of
p-form field strengths, this is not possible by perturbing an
isomety. For a second reason, consider the continuous
isometries of the metric: $SU(2)\times SU(2)\times U(1)$. The $U(1)$
isometry is just the M theory circle. If we would like the fluctuation
to reduce to a fluctuation about the corresponding IIA background,
then we should not perturb about this direction. The remaining $U(1)$
isometries are contained within $SU(2)$s. But ${\mathcal{N}}=1$ SYM
does not have a global $SU(2)$ symmetry and in particular, the
anomalous chiral $U(1)$ is not contained within an $SU(2)$.

A check of the idea of perturbing a non-isometric angle is as
follows. Consider a similar background expected to be dual to a
field theory without a chiral anomaly, and see whether the
symmetry breaking effect does not occur. Supersymmetric, purely
gravitational backgrounds of M theory can be systematically
constructed from considering D6-branes wrapping calibrated cycles
in IIA backgrounds \cite{Gomis:2001vk}. For example, N D6-branes
wrapping an $S^2$ inside a nonsingular $K3$ manifold. The theory
living on these D-branes is five dimensional $SU(N)$ SYM theory
with eight real supercharges. This theory will not have an
anomalous chiral symmetry, because it is five dimensional. The
configuration is described in eleven dimensions by M theory on the
resolved conifold with a singular $\Z_N$ quotient
\cite{Edelstein:2001pu}. The resolved conifold metric is
asymptotically conical with base $T^{1,1}$. The $T^{1,1}$ contains
an angle, $\psi'$, that is very similar to the $\psi$ angle that
we have just considered. In fact, if one reduces the $G_2$ metrics
to IIA, then we also obtain an asymptotically conical geometry
with base $T^{1,1}$, and with $\psi$ the same angle in the
$T^{1,1}$ as $\psi'$. An important difference between the two
cases is that in the $G_2$ metrics, the direction $\pa_\psi$ is
not an isometry in the interior of the metric, whilst in the
resolved conifold, $\pa_{\psi'}$ remains an isometry everywhere.
Thus we find that the existence of an interesting non-isometric
direction is correlated with the existence of the chiral anomaly.

The previous paragraph gives another reason to take $\pa_{\psi}$
as the massive isometry. The shift $\psi \to \psi + \epsilon$ of
the angle $\psi$ in $T^{1,1}$ is generically dual, in string
theory conifold backgrounds, to the field theory R-symmetry
\cite{Klebanov:1998hh}.

\section{Consistency and inconsistency of fluctuations}
\label{sec:testI}

{\it In this section, we show that the fluctation about the $G_2$
backgrounds that we have just considered is asymptotically
consistent in general and fully consistent in the $\bmet$ and
$\cmet$ cases.}

\subsection{Equations of motion and consistency when $g(r)=1$}

Let us first calculate the equations of motion corresponding to
the action we just derived (\ref{eq:Waction}). It is convenient to
define the two functions
\bea\label{eq:PandQ}
P(r)^2 & = & \frac{(g^2-1)^2 [2abf
+ c(a^2-b^2)]^2}{16 a^2 b^2}+ c^2 \, ,\nonumber\\
Q(r)^2 & = & \frac{[a^2-b^2]^2}{2a^2 b^2} \, ,
\eea
So the action (\ref{eq:Waction}) is written
\be
S \propto - \int d^4x\,dr \sqrt{g_7(r)}
\left[ \frac{P(r)^2}{4}  F_{a b}
  F^{a b} + \frac{Q(r)^2}{2} W_a W^a \right] .
\ee
The equations of motion that follow from this action are
\bea\label{eq:eqnsofmotion}
\6_j F^{jr} & = & \frac{Q^2}{P^2} W^r \nonumber\\
\6_r F^{ri}+ \6_j F^{ji} + \frac{\pa_r (\sqrt{g_7} P^2)}{\sqrt{g_7} P^2} F^{ri} & = &
\frac{Q^2}{P^2} W^i\,.
\eea
We use indices $i,j \ldots$ to run over the coordinates $t,x,y,z$,
whilst as before $a,b \ldots$ run over $t,x,y,z,r$. Now consider
$\6_r$ of the first equation and subtract $\6_i$ of the second.
The equation obtained is of type well known for massive vector
fields
\be
\label{divW}
\6_a W^a + \left[ \frac{\pa_r (\sqrt{g_7} P^2)}{\sqrt{g_7}P^2}+ \frac{P^2}{Q^2}
 \pa_r \left(\frac{Q^2}{P^2}\right)\right] W_r =0\,.
\ee

We now need to see whether these equations solve the full eleven
dimensional equations of motion. This is necessary in order for the
fluctuations, which are eleven dimensional, to be on shell and hence
physical.

As an illustration of when this works, consider first the case of
the allowed truncation to $g=1$. This corresponds to considering
the $\bmet$ and $\cmet$ metrics. The equations of motion
(\ref{eq:eqnsofmotion}) in this case follow from the action
(\ref{eq:gisoneaction}) and the first order equations
(\ref{eq:firstorder})
\bea\label{eq:gisoneqns}
& \6_j F^{jr} &  =  \frac{(a^2-b^2)^2}{2 a^2 b^2 c^2} W^r \nonumber\\
& \6_r F^{ri} & + \6_j F^{ji} - \frac{cf (b^2-a^2) + 2 a b (3 a^2 + 3 b^2 -
c^2)}{4 a^2 b^2 c} F^{ri} =
\frac{(a^2-b^2)^2}{2 a^2 b^2 c^2} W^i\,,
\eea
while the divergence expression (\ref{divW}) becomes
\be
\label{gisonedivW}
\6_a W^a - \frac{cf [(a^4+b^4)+6a^2 b^2] + 2
ab[c^2(a^2-b^2)+b^4-a^4]}{4 a^2 b^2 c (a^2-b^2)} W_r =0\,.
\ee

The equations of motion satisfied by the background metric in eleven dimensions
(\ref{eq:g2metric}) are
\be
R_{\bar{u}\bar{v}} = 0 ,
\ee
where indices $\bar{u},\bar{v}$ run over the eleven tangent space coordinates.
After including the fluctuation (\ref{eq:fluctuation}) in the
metric, the Ricci tensor acquires the following nonzero
components. It is important to note firstly that we are working to
linear order in the fluctuation only and secondly that we work in
a gauge where $\lambda=0$, in which therefore $A = W$.
\be
\d R_{\ih\bar{10}} = \frac{c}{2} \left[ - \6_r F^{ri} - \6_j F^{ji} +
\frac{cf (b^2-a^2) + 2 a b (3 a^2 + 3 b^2 -
c^2)}{4 a^2 b^2 c} F^{ri} +
\frac{(a^2-b^2)^2}{2 a^2 b^2 c^2} W^i \right] \,, \label{eq:line1}
\ee

\be
 \d R_{\bar{5}\bar{10}} = \frac{c}{2} \left[ - \6_j F^{jr} +
 \frac{(a^2-b^2)^2}{2 a^2 b^2 c^2} W^r \right]
\label{eq:line2} \,,
\ee

\be
\d R_{\bar{6}\bar{9}} = - \d R_{\bar{8}\bar{7}} = \frac{(a^2-b^2)}{4 a
b} \left[ \6_a W^a - \frac{cf [(a^4+b^4)+6a^2 b^2] + 2
ab[c^2(a^2-b^2)+b^4-a^4]}{4 a^2 b^2 c (a^2-b^2)} W_r \right] \,.
\label{eq:line3}
\ee
The bars denote tangent space coordinates. The numbers correspond
with the order that the vielbeins appear in the metric
(\ref{eq:g2metric}). Thus for example the $\bar{10}$ direction
corresponds to the $u g^5$ vielbein. The equations for the radial
functions (\ref{eq:firstorder}) have been used.

It is immediate that the three equations of motion
(\ref{eq:gisoneqns}) and (\ref{gisonedivW}) imply the vanishing of
$\d R_{\bar{u}\bar{v}}$. Therefore the reduction is consistent to
first order in the fluctuation. At various stages we used the
first order equations (\ref{eq:firstorder}). We needed
specifically the first order equations for $G_2$ holonomy. The
second order equations for Ricci flatness of the metric
(\ref{eq:g2metric}) would not have been sufficient, although they
would have been sufficient to derive the action
(\ref{eq:Waction}). Thus, special holonomy plays an important role
in the consistency of the fluctuation.

\subsection{Asymptotic consistency for general $G_2$ metrics}
\label{sec:consistency}

Now let us consider the more general $G_2$ holonomy metrics, which
do not have $g(r)=1$. Now things do not work so clearly. To show
that the perturbation is in fact inconsistent in general, we will
substitute the equations of motion (\ref{eq:eqnsofmotion}) and
(\ref{divW}) into the fluctuation of the Ricci tensor $\d
R_{\bar{u}\bar{v}}$ and find that it is not zero, and that the
remaining nonzero terms cannot be set to zero by the equations of
motion.

Concretely, after using the equations of motion, the fluctuation
of the Ricci tensor has the following nonzero components
\be\label{eq:inconsistent}
\d R_{\ih\bar{11}} = (g^2-1)\left(\frac{(b^2-a^2)^2 [c (a^2-b^2) + 2
abf]}{\sqrt{C(r)}}  W_\ih + \frac{B(r)}{\sqrt{C(r)}} F_{\ih r}  \right) \,,
\ee
\be
\d R_{\bar{5}\bar{11}} = (g^2-1) \frac{(b^2-a^2)^2 [c (a^2-b^2) + 2 abf]}{\sqrt{C(r)}}  W^r \,,
\ee
where $B(r),C(r)$ are complicated functions of the radial
functions $a,b,c,f,g$.  Their precise form does not seem to be
illuminating, so we don't include them here. The only common
factor amongst these terms is $g^2-1$. These terms cannot be set
to zero using the reduced equations of motion. Therefore, the
fluctuation is inconsistent unless $g^2=1$.

At first sight, this seems problematic for the $\dmet$ and $\ctmet$
metrics which have $g \neq 1$. However, all the families of $G_2$
holonomy metrics described by (\ref{eq:g2metric}) have the same
asymptotics as $r \to \infty$. In particular $g^2 \to 1$
asymptotically, which suggests a notion of asymptotic consistency.

In fact, whilst all the metrics have the same leading order
asymptotics, there are two possible forms at subleading orders.
The first is that of the $\bmet$ and $\cmet$ metrics and is
\bea\label{eq:expand1}
a(r) & \sim & \frac{r}{\sqrt{12}} + \frac{\sqrt{3}}{2} \left( q - \frac{R}{2} \right)
+ \frac{3 \sqrt{3} R^2}{4 r} +  \cdots\,, \nonumber \\
b(r) & \sim & - \frac{r}{\sqrt{12}} - \frac{\sqrt{3}}{2} \left( q + \frac{R}{2}
\right) - \frac{3 \sqrt{3} R^2}{4 r} + \cdots \,, \nonumber \\
c(r) & \sim & \frac{r}{3} + q + \frac{3 R^2}{r} + \cdots \,, \nonumber \\
f(r) & \sim & R - \frac{9 R^3}{r^2} + \cdots \,, \nonumber \\
g(r) & = & 1 \,,
\eea
where $R,q$ are free constants. These solutions have $n=m$, where $n$
and $m$ are the integration constants of (\ref{eq:conserved}).

The second form of subleading asymptotics, which
is that of the $\dmet$ and $\ctmet$ metrics, is
\bea\label{eq:expand2}
a(r) & \sim & \frac{r}{\sqrt{12}} + \frac{\sqrt{3}}{2} \left( q - \frac{R}{2} \right)
+ \frac{\sqrt{3} \left( 6 R^3 + m \right) }{8 R r} +  \cdots\,, \nonumber \\
b(r) & \sim & - \frac{r}{\sqrt{12}} - \frac{\sqrt{3}}{2} \left( q + \frac{R}{2}
\right) - \frac{\sqrt{3} \left( 6 R^3 + m \right) }{8 R r} + \cdots \,, \nonumber \\
c(r) & \sim & \frac{r}{3} + q + \frac{3 R^2}{r} + \cdots \,, \nonumber \\
f(r) & \sim & R + \frac{3m - 9 R^3}{r^2} + \cdots \,, \nonumber \\
g(r) & \sim & 1 - \frac{3m}{2 R r^2} + \cdots \,,
\eea
where $m$ is a new constant. For the $\dmet$ metrics, $m$ is just
the first integration constant in (\ref{eq:conserved}). The
$\dmet$ metrics further have $n=0$, where $n$ is the other
integration constant of (\ref{eq:conserved}).

With these expansions we may study the question of consistency at
large $r$. If one calculates the Ricci tensor at large $r$ in tangent space
coordinates one finds the following, in a hopefully obvious notion,
\bea
\d R_{\ih\bar{10}} & \sim & \O (r) \pa F + \O (1) F + \O
( r^{-3} ) W \,, \nonumber \\
\d R_{\bar{5}\bar{10}} & \sim & \O ( r ) \pa F +
\O ( r^{-3} ) W \,, \nonumber \\
\d R_{\ih \bar{11}} & \sim & \O ( r^{-3} ) F +
\O ( r^{-6} ) W \,, \nonumber \\
\d R_{\bar{5}\bar{11}} & \sim &\O ( r^{-6} ) W \,, \nonumber
\\
\d R_{\bar{6} \bar{9}} & \sim & \d R_{\bar{7} \bar{8}} \sim \O
( r^{-1} ) \pa W + \O ( r^{-2} ) W \,.
\eea
These results show that the inconsistent components, $\d R_{\ih \bar{11}},
\d R_{\bar{5}\bar{11}}$ have coefficients of $W_{\ih}$ and $F_{\ih \bar{r}}$
that are suppressed by three powers of $r$ compared with the
consistent terms. Thus at large $r$ the fluctuation solves the
full eleven dimensional equations of motion to a good
approximation.

Below we will suggest that, as is common in gauge-gravity
dualities, large radius corresponds to the UV of the dual field
theory. The physical interpretation of the asymptotic consistency
is that we have found the correct gravitational description of the
field theory symmetry current in the UV but that a more
complicated description is required in the IR. This is not
necessarily surprising, given that as well as being anomalously
broken to $\Z_{2N}$, the chiral symmetry is further spontaneously
broken in the IR to $\Z_{2}$.

The asymptotic consistency of the fluctuation is an important test
of our proposal for spontaneous symmetry breaking. We will see in
the last section below that such simple vector fluctuations about
a generic non-isometric direction are not asymptotically
consistent. Thus the $\pa_\psi$ direction is special in this
regard.

\section{Comparison with generic five dimensional results}

{\it In this section we dimensionally reduce the action for
our perturbation to five dimensions. We obtain a massive vector field
with a mass that is asymptotically precisely that predicted on general
grounds for vector fields dual to the chiral current.}

\subsection{A five dimensional prediction}

In studying holographic renormalisation group flows, a general
prediction was made for the mass of a vector field dual to a
broken chiral symmetry \cite{Bianchi:2000sm}. Five dimensional
backgrounds of the form
\be\label{5dmetric}
ds^2_5 \equiv g_{ab} dx^a dx^b = dq^2 + e^{2T(q)} dx^2_4,
\ee
were considered. It was argued that the transverse part of the
vector field, V, associated with the chiral symmetry, would have
an action, with the metric in the Einstein frame, given by
\be\label{eq:genaction}
S = - \int d^4 x \,dq \sqrt{-g} \left[\frac{1}{4} G_{ab} G^{ab} -
\frac{1}{2} 2 \frac{\pa^2 T}{\pa q^2}  V_a V^a \right]\,,
\ee
where $G = d V$. In particular, the mass is $m^2 = - 2 \pa_q \pa_q
T$. The reason one only studies the transverse part, with $V^q =
\nabla_a V^a = 0$, is that in general one does a gauge field
redefinition to obtain the gauge-kinetic term with canonical
normalisation. The action for the $V_q$ and divergence components
become more complicated because of explicit $q$-dependence in the
action.

We would thus like to know whether a dimensionally reduced version
of our perturbation to five dimensions may be put into this
form.

\subsection{Dimensional reduction of the $G_2$ fluctuation}

Start with the action (\ref{eq:Waction}) written as
\be
S = - \int d^4 x \,dr \sqrt{-g_5} \sqrt{g_6} \left[\frac{1}{4}
P(r)^2 F_{ab} F^{ab} + \frac{1}{2} Q(r)^2 W_a W^a \right]\, .
\ee
The five dimensional metric is just
\be
ds^2_5 = dr^2 + dx^2_4 ,
\ee
and the determinant $g_6$ is obtained by integrating out the
angles of the $G_2$ holonomy metric from the determinant
(\ref{eq:determinant}) where we have now included the numerical
factor from integrating out the angles:
\be
g_6 = \frac{2^{10}\pi^4}{N} a^2 b^2 c^2 g^4
\left[2abf(1+g^2) + c(b^2-a^2)(1-g^2)
\right]^2 .
\ee
We only work with the transverse components, so again $W^r =
\del_a W^a = 0$. Recall that transverse and scalar components do
not mix at a linearised level, so this truncation is consistent.
In order to compare with the result we quoted in the previous
subsection, we need to be in the Einstein frame. Thus we define
\be\label{eq:defphi}
e^{-6\phi(r)} = g_6\,,
\ee
and define a rescaled metric by
\be\label{eq:Einsteinframe}
d\tilde{s}_5^2 = e^{-2\phi(r)} ds^2_5 \,.
\ee
In terms of this metric, the action now becomes
\be
S = - \int d^4 x \,dr \sqrt{-\tilde g_5} \left[ \frac{1}{4} P(r)^2
  F_{\tilde g}^2 e^{-2\phi} + \frac{1}{2} Q(r)^2 W_{\tilde g}^2
  \right] \,,
\ee
where the $\tilde g$ subscript denotes that the indices are
contracted with the rescaled metric, $\tilde g_5$.  We must now
perform a field redefinition to obtain a canonically normalised
kinetic term for the gauge field. Defining a new field, $V_a$, by
\be
W_a = \frac{e^\phi}{P} V_a\,,
\ee
the action takes the form
\be
S = - \int d^4 x \,dr \sqrt{-\tilde g_5} \left[ \frac{1}{4}
  G_{\tilde g}^2 + \frac{1}{2} m(r)^2 V_{\tilde g}^2
  \right] \,,
\ee
where again $G = d V$ and the mass is given by
\be
\label{eq:masssquared}
m(r)^2 = e^{2\phi} \left[ - \phi'' + 2 (\phi')^2 - 3
\frac{P'\phi'}{P} + \frac{P''}{P}+ \frac{Q^2}{P^2} \right]\,,
\ee
where the primes denote differentiation with respect to $r$. The
action is now in the same form as the action (\ref{eq:genaction})
with which we wish to compare the mass term.

We can use the asymptotic expressions for the radial functions
(\ref{eq:expand1}) and (\ref{eq:expand2}) to find that
asymptotically (\ref{eq:masssquared}) becomes
\be
m(r)^2 = e^{2\phi} \left[
\frac{80}{9r^2}+{\mathcal{O}}\left(\frac{R^2}{r^4}\right)
\right] \,.
\ee
Now compare this with the prediction from the previous subsection.
Note that from comparing (\ref{5dmetric}) and
(\ref{eq:Einsteinframe}) we see that $dq = e^{-\phi(r)} dr$ and
$T(q) = -\phi(r)$. Thus:
\be\label{eq:prediction}
m^2_{\text{prediction}} = - 2 \frac{\pa^2 T}{\pa q^2} = 2
e^{2\phi} \left[\phi'' + (\phi')^2 \right]
\, ,
\ee
which is asymptotically
\be
m^2_{\text{prediction}} = e^{2\phi} \left[ \frac{80}{9r^2}
  +{\mathcal{O}}\left(\frac{R^2}{r^4}\right) \right]  \,.
\ee
Thus we have found exact agreement to leading order with the
general prediction of \cite{Bianchi:2000sm} The agreement does not
continue to subleading orders.

The leading order matching we have just described does not depend
on the eleven dimensional mass, $Q^2/P^2$, which turns out to be
subleading in (\ref{eq:masssquared}). It is important to emphasise
that the five dimensional mass we are considering here has a
different character to the mass in previous sections. Even an
isometry will generically have a mass term if one rescales the
gauge field to get canonical normalisation. The nontrivial point
of the previous section is that when one has a massive isometry,
the mass generation is from spontaneous symmetry breaking. The
canonical normalisation was not the appropriate normalisation to
use in that context.

\subsection{Generic agreement to leading order}

The agreement we have just found supports the setup we are
presenting. Unfortunately, the particular direction we are perturbing
is not the unique direction that would have resulted in an agreement. More
concretely, from comparing (\ref{eq:masssquared}) and
(\ref{eq:prediction}) one sees that the following conditions are
sufficient to imply agreement to leading order as $r \to \infty$:
\bea\label{eq:agreement}
\frac{P''}{P} \, \text{ and } \, \frac{Q^2}{P^2} & \ll & \phi'' \,, \nonumber \\
P & \propto & \frac{1}{\phi'} .
\eea
From (\ref{eq:defphi}) we see that whenever the determinant $g_6$
is polynomial to leading order as $r \to \infty$, then $1/\phi'
\sim r$. This is the case in all the $G_2$ metrics we have been
considering. Now recall that the $G_2$ metrics are asymptotically
locally conical, and that $P(r)$ is just the radial function in
the metric in front of the angle we perturb. Thus, for a generic
angle in the metric, except $\pa_\gamma$ which stabilises at
infinity, we will have $P(r) \sim r$. The second condition in
(\ref{eq:agreement}) is therefore satisfied. This also implies
that the first half of the first condition, $P''/P \ll
\phi''$ is generically satisfied. The remaining condition $Q^2/P^2 \ll
\phi''$ would need to be checked for every consistent
perturbation, but it will certainly be satisfied by isometries,
which have no mass in eleven dimensions: $Q_{\text{isometry}}=0$.

Therefore, if we had perturbed any of the isometries in the $G_2$
holonomy metric, besides the M-theory $U(1)$, we would also have
found agreement to leading order. At the end of section
\ref{sec:g2fluctuation} we gave a few reasons however for why isometries
of the metric were unlikely to be relevant for the dual description of
the chiral anomaly.

The conclusion of this section seems to be that although one
obtains encouraging results from a five dimensional perspective,
the dimensionally reduced picture needs to be better understood.
In particular, it would be nice to embed the fluctuation into a
five dimensional supergravity theory. This would enable one to see
whether the vector fluctuation is in the same supersymmetry
multiplet as the metric, as is usually the case for R-symmetry
gauge fields. This is because the R-symmetry current and the
energy momentum tensor of the field theory are in the same anomaly
multiplet. One route towards this could be via eight dimensional
supergravities, which have been used to construct $G_2$ metrics
\cite{Edelstein:2001pu,Hernandez:2002fb,Edelstein:2002zy} .

\subsection{Agreement for the Maldacena-N\'u\~{n}ez background}

We can easily apply the methods we have just developed to the
fluctuation about the MN background that we considered in section
\ref{sec:SSBMN}. However, first we should note that at large $r$,
the D5-brane solutions we described has diverging dilaton coupling
(\ref{dil}). Therefore, if we wish not to deal with large stringy
corrections, we should use the S dual solution describing
NS5-branes. In the Einstein frame solution (\ref{MNUV}) this
simply corresponds to letting $\Phi \to - \Phi$ in the metric.
Further, the dilaton now has asymptotic behaviour
\be
e^{2\Phi}=e^{2\Phi_0}\frac{4\sqrt{r}}{e^{2r}}.
\label{dilNS}
\ee
The ten dimensional action for the fluctuation is now given by
\be
S\propto - \int d^{10}x \sqrt{g_{10}}
\left[\frac{N}{16} e^{-\Phi/2} |F_2|^2 + \frac{3(16r^2 +1)}{16r^2} e^{2\Phi}
W^2\right] .
\label{actionNS}
\ee
This action may be reduced to five dimensions and the metric
rescaled to be in the five dimensional Einstein frame in
essentially the same way as in the previous two subsections.

The generic prediction and the dimensionally reduced masses agree
to leading order, and are given by
\be
m(r)^2 \propto r^{2/3} e^{-4\Phi/3} (\Phi')^2 + \cdots \propto
r^{1/3} e^{r/3} .
\ee
If one had not S dualised, then the leading coefficient of the
masses would not have agreed, showing that stringy corrections can
upset the matching. This is in contrast to the $G_2$ holonomy
backgrounds, where one can use the same $\dmet$ metric at large
and small $r$.

\section{Wrapped D5 and D6-branes, and decoupling}

\textit{The $G_2$ metrics we are studying are related to wrapped
D6-branes. This section considers the decoupling of Kaluza-Klein
modes and gravity from the dual field theory. Comparisons with the
Maldacena-N\'u\~{n}ez solution of wrapped D5-branes are made. We
discuss the breaking $U(1) \to \Z_{2N}$}.

\vspace{0.5cm}

\noindent The $G_2$ manifolds we are discussing are M theory lifts of
supersymmetric configurations of N D6-branes wrapped on a
three-cycle in a Calabi-Yau manifold
\cite{Vafa:2000wi,Atiyah:2000zz,Acharya:2000gb}. The M theory circle
is the $U(1)$ isometry $\pa_\gamma$. The low energy theory living on
the wrapped D6-branes is ${\mathcal{N}}=1$ SYM. Although the
first of the new $G_2$ metrics constructed in this context were
the $\bmet$ metrics \cite{Brandhuber:2001yi}, these metrics have
an ADE singularity at the origin and do not describe the full
back-reaction of the branes on the geometry.

The subsequently discovered $\dmet$
\cite{Cvetic:2001ih,Brandhuber:2001kq} metrics are not only
well-behaved at the origin, but have the topology expected from
studies of the M theory flop
\cite{Atiyah:2000zz,Acharya:2000gb,Acharya:2001hq,Acharya:2001dz}
to describe the confining IR of ${\mathcal{N}}=1$ gauge theory. Recall
that the topology of the $\dmet$ metrics is $S^3/\Z_N\times \R^4$. At
large radial direction, the $\dmet$ metrics have the same
asymptotics as the $\bmet$ metrics, see equations
(\ref{eq:expand1}) and (\ref{eq:expand2}) above, and thus should
also capture some of the UV physics, as we are arguing in this
paper.

The last two paragraphs suggest that the $\dmet$ metrics are the
dual geometries to wrapped D6-branes in Type IIA string theory, in
the same way that the Maldacena-N\'u\~{n}ez solution is the dual
geometry to wrapped D5-branes in type IIB theory.
The IIA reduction of the $\dmet$ metrics is the resolved conifold with
$N$ units of RR flux through the noncollapsed $S^2$.
Within the
geometric transition perspective introduced in \cite{Vafa:2000wi},
this expectation of duality is very natural.

The following discussion about decoupling is applicable to both the
wrapped D5 and D6 configurations. For most of the time, we shall
discuss only the D6 case explicitly.

There are two issues of decoupling when considering wrapped
branes. One is the decoupling of Kaluza-Klein modes on the wrapped
cycle and the other is the decoupling of gravity from field theory in
the `field theory limit'.

\subsection{Kaluza-Klein modes}

First consider the Kaluza-Klein modes. The fact that the wrapping
of the branes involves twisting suggests that the somewhat na\"ive
analysis we are about to present may not be the whole story. We
will say few words about this below. For this discussion only, we
work explicitly with factors of $\alpha^{\prime}$.

The mass scale associated with the Kaluza-Klein modes should be
\be
\Lambda_{\text{KK}}^3
\equiv M_{\text{KK}}^3 \sim \frac{1}{\text{Vol}S^3} ,
\ee
where $S^3$ is the cycle in the Calabi-Yau wrapped by the branes.
There is no a priori reason to identify this cycle with any
particular cycle in the backreacted $G_2$ holonomy geometry.

We would like to compare this scale with the characteristic mass
scale of the field theory generated through dimensional
transmutation. This is defined to be the energy scale at which the
one loop coupling diverges. Up to the Kaluza-Klein mass scale, the
theory is effectively four dimensional and the coupling constant
will follow the ${\mathcal{N}}=1$ beta function. In particular, we
may use the logarithmic one loop running of the effective coupling
at the energy scale $\Lambda_{\text{KK}}$ to obtain the following
expression for the super Yang-Mills energy scale
\be\label{eq:SYMscale}
\Lambda_{\text{SYM}}^3 \sim \Lambda_{KK}^3
e^{\frac{- 8 \pi^2}{g^2_{\text{eff,4}}(\Lambda_{KK})}} .
\ee
Now, at the energy scale $\Lambda_{KK}$ we should match the
dimensionless four dimensional 't Hooft effective coupling with
the dimensionless seven dimensional effective coupling
$g^2_{\text{eff,4}}(\Lambda_{KK}) =
g^2_{\text{eff,7}}(\Lambda_{KK})$. Further, on dimensional
grounds, we know how to relate the dimensionless effective
coupling to the dimensionful Yang-Mills coupling that appears in
the action $g^2_{\text{eff,7}}(\mu) = \mu^3 N g^2_{\text{YM,7}}$.
In particular, this implies $g^2_{\text{eff,7}}(\Lambda_{KK}) = N
g^2_{\text{YM,7}}/\text{Vol}S^3$. Finally, we can relate the
Yang-Mills coupling to string theory quantities and hence to the M
theory Planck length in the standard way \cite{Itzhaki:1998dd}:
$g^2_{\text{YM,7}} = (2\pi)^4 g_s \alpha^{\prime 3/2} = (2\pi)^4
l_p^3$. Combining all these statements allows us to re-express
(\ref{eq:SYMscale}) as
\be
\Lambda_{\text{SYM}}^3 \sim \Lambda_{KK}^3
e^{\frac{- \text{Vol}S^3}{N 2 \pi^2 l_p^3}} .
\ee
To decouple the Kaluza-Klein modes, we would like to take
\be
\Lambda_{\text{SYM}} \ll \Lambda_{\text{KK}} ,
\ee
which requires
\be\label{eq:decouple}
\text{Vol}S^3 \gg N 2 \pi^2 l_p^3 .
\ee
This is always possible in principle, the question is whether it
is compatible with the regime of validity of the gravity dual.

On general grounds, we should not expect a perturbative regime in
field theory to overlap with a regime in which a supergravity
description is valid, else the conjectured duality would be
falsified by the fact that the two theories are manifestly
different. An example of this is the well-studied case of near
horizon geometries of flat D-branes, where the scalar curvature
in the string frame satisfies \cite{Itzhaki:1998dd}
\be\label{eq:coupling}
\alpha^{\prime} R_{\text{str}} \sim \frac{1}{g_{\text{eff}}} ,
\ee
so clearly perturbative field theory and small gravitational
curvatures are incompatible. In general the dependence of
$g_{\text{eff}}$ on the energy scale, $\mu$, translates into a
radial, $r$, dependence of $\alpha^{\prime} R$. Typically, small
$r$ will be the IR of the field theory, and large $r$ will be the
UV.

It seems plausible that to have a valid gravitational description,
the dual effective coupling should be large. The minimum of the
effective coupling is at $\Lambda_{\text{KK}}$ because above this
scale it will increase according to the seven dimensional relation
$g^2_{\text{eff,7}}(\mu) = \mu^3 N g^2_{\text{YM,7}}$, and below
this scale it will increase due to the logarithmic running of the
four dimensional coupling. This is illustrated in the following
figure.

\begin{figure}[h]
\begin{center}
\epsfig{file=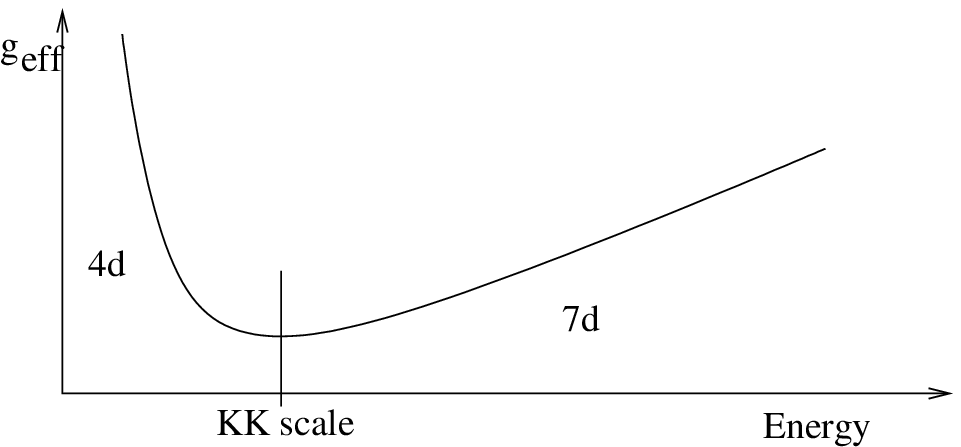,width=8cm}

\noindent {\bf Figure 2:} The running of the effective coupling.
\end{center}
\end{figure}

We have just seen that $g^2_{\text{eff,4}}(\Lambda_{KK}) =
\frac{N 16 \pi^4 l_p^3}{\text{Vol}S^3}$. Therefore the regime in which
the effective coupling is large, $\text{Vol}S^3 \ll N 16 \pi^4
l_p^3$, has no overlap with the decoupling regime
(\ref{eq:decouple}). This is the standard problem of strong
coupling gauge theory duals: in order to obtain the dual, one
introduces a new scale, in this case the Kaluza-Klein scale, which
then cannot be decoupled from the field theory scale. Note that in
our argument we have not made any assumptions relating certain
cycles within the gravitational background with the initial cycle
the branes wrap, as is often done in discussions of the
Maldacena-N\'u\~{n}ez background, for example
\cite{Bigazzi:2003ui,Loewy:2001pq}. Instead we have made the
assumption that the gravitational background should not contain a
regime dual to a weakly coupled field theory.

As we noted above, the analysis we have just presented may be too
na\"{i}ve. In particular, we have not taken into account the fact
that the theory living on wrapped D-branes is partially twisted
\cite{Bershadsky:1995qy} in order to be supersymmetric. The
degrees of freedom along the wrapped directions are topological.
One might expect this fact to modify the role of the Kaluza-Klein
scale in the theory and the running of the coupling above the
Kaluza-Klein scale.

So far we have discussed general expectations from the
renormalisation group flow in field theory. Now we turn to the
other side of the duality and calculate curvatures and string
couplings for the gravity backgrounds. It is not difficult to
explicitly calculate the Ricci scalar and dilaton for the string
frame IIA backgrounds resulting from dimensional reduction of the
$\dmet$ metrics \cite{Brandhuber:2001kq}. One finds that the Ricci
scalar decreases away from the origin. As we are now discussing
the background itself, we return to units with
$\alpha^{\prime}=1$. One finds
\be
R_{\text{str, r=0}} \sim {\mathcal{O}}\left(\frac{N}{L^2}\right) ,
\ee
where here $L$ is the radius of the $S^2$ that does not collapse
at the origin in the IIA metrics, and carries $N$ units of
Ramond-Ramond flux. On the other hand, the dilaton increases away
from the origin. For a typical value of the nontrivial parameter
of the $\dmet$ metrics, i.e. not the scaling parameter,
(\ref{eq:expand2}), the dilaton coupling remains the same order of
magnitude. That is
\be
g_s \equiv e^{\Phi} \sim {\mathcal{O}}\left(
\frac{L}{N^{3/2}}
\right) .
\ee
The last two formulae imply that
\be
R_{\text{str, r=0}} \sim \frac{1}{g_s^2 N^2} ,
\ee
which might be compared with the result for the
Maldacena-N\'u\~{n}ez solution: $R_{\text{str}} \sim 1/g_s N$.

Thus small curvature requires $g_s N \gg 1$. This will be
consistent with small string coupling if we take $N$ large. Unlike
the Maldacena-N\'u\~{n}ez case, the dilaton remains finite at
infinity, and so the lift to M theory is not always forced.

The $G_2$ manifold itself is of course Ricci flat, but one can
calculate the Riemann squared curvature invariant. This is seen to
decrease to zero away from the origin. At the origin one has
\be
\left. R^{a b c d} R_{a b c d}\right|_{\dmet, r=0}
\sim {\mathcal{O}}\left(\frac{1}{\tilde{L}^4}\right) ,
\ee
where $\tilde{L}$ is the radius of the noncollapsed $S^3$ in the
$\dmet$ metric. We can then safely take $\tilde{L}$ large to avoid
large M-theoretic corrections to the background.

\subsection{Gravitational modes}

The decoupling limit of D6-branes is well-known to be problematic
\cite{Itzhaki:1998dd}. The $\alpha^{\prime} \to 0$ limit is taken
with $g_{\text{YM,7}}$ held fixed. As we have just recalled,
$g^2_{\text{YM,7}} = (2\pi)^4 g_s \alpha^{\prime 3/2} = (2\pi)^4
l_p^3$, which implies a finite eleven dimensional Planck length
and hence the non-decoupling of gravity. Further, the
characteristic mass scale of gravitational contamination in the
field theory is $1/l_p$. The discussion of the previous subsection
shows that we cannot decouple $\Lambda_{KK}$ and $1/l_p$ within
the regime of gravitational description. Therefore the
gravitational scale also cannot be decoupled from $\Lambda_{SYM}$.

Two features that occur in the M theory lift of flat D6 branes,
and which are related to the non-decoupling of gravity
\cite{Itzhaki:1998dd} also occur in the $G_2$ metrics. Firstly,
$R^{abcd}R_{abcd}$ goes to zero asymptotically. Secondly, the
curvatures of the metric do not depend on $N$ at all.

The analogous problem in the D5-branes wrapping an $S^2$ of the
Maldacena-N\'u\~{n}ez solution is the decoupling of the little
string scale. However, the D6-brane case is worse because the
non-decoupling of gravity means that the field does not have UV
completion other than M theory itself.

It is remarkable that despite such problems with decoupling,
dualities involving wrapped branes are able to reproduce features
of pure ${\mathcal{N}} = 1$ Super Yang Mills theories. Most
features are qualitative or of a topological character
\cite{Bigazzi:2003ui}, so it is particularly mysterious that the
perturbative beta function for ${\mathcal{N}} = 1$ Super Yang
Mills was recently calculated from the Maldacena-N\'u\~{n}ez
solution \cite{DiVecchia:2002ks,Bertolini:2002yr,Apreda:2001qb}.
Perhaps the success is related to the fact, as we mentioned
previously, that the degrees of freedom in the wrapped directions
of the brane are described by a topological field theory.

Another quantitative matching that has been possible in other
other dualities is the manifestation of the anomalous breaking
$U(1)_R \to \Z_{2N}$ in the gravity side of the duality. We turn
to this point next, and see that whilst the MN case works very
straightforwardly, the $G_2$ case does not. This may be related to
the different nature of the non-decoupling in the two models,
little string modes as opposed to full M theory modes.

\subsection{$U(1) \to \Z_{2N}$ in the Maldacena-N\'u\~{n}ez background}
\label{sec:z2n}

Let us review how this mechanism works in the MN solution
\cite{Bigazzi:2003ui}. Recall from the introduction that the
metric (\ref{MNUV}) has a symmetry under
\be\label{iso1}
\psi\to\psi+\eps.
\ee
The metric is invariant under this constant shift. It will be a
symmetry of the background if the shift (\ref{iso1}) is also a gauge
transformation of the RR potential.

In fact, the RR potential (\ref{RRpot}) transforms as
\be\label{C2trans}
C_2 \to C_2 + \frac{\eps}{4} N \,\, \omega_2
\ee
under (\ref{iso}). Classically, this is a gauge transformation
because $d\omega_2 = 0$. However, there are quantum mechanical
complications because although $\omega$ is closed, it is not
exact. This is possible because of the topology at infinity
$H^2(S^2\times S^3,\Z)=\Z$. To see the effect of this, consider a
probe instantonic D1-brane that is coupled to $C_2$ with the
following contribution to the partition function
\be\la{probe}
Z_{D1}[C_2] \sim e^{\frac{i}{2\pi}\int_{S^2}C_2} ,
\ee
where $S^2$ has $\omega_2$ as volume form. $Z_{D1}$ is required
to be invariant under the gauge transformations, $C_2\to C_2 +
\lambda_2.$ While this invariance is trivial for an exact
$\lambda_2$, one should also require the invariance under `large'
gauge transformations where $\lambda_2$ is proportional to the
volume form on $S^2$, that is $\lambda_2 = c \,\,
\omega_2$. Then invariance of (\ref{probe}) fixes $c$ to be an
arbitrary integer multiple of $\frac{\pi}{2}$. Thus we conclude
that as long as the shift of $C_2$ under the chiral transformation
(\ref{C2trans}) is equivalent to a ``large'' gauge transformation
$C_2 \to C_2 + n \frac{\pi}{2} \omega_2$, the chiral
transformation will be a symmetry of the whole background. This
condition fixes
\be
\eps=\frac{2 n \pi}{N} ,
\ee
where $n$ is an arbitrary integer. Since $\eps$ is defined modulo
$4\pi$ we see that there are $2N$ discrete values of the gauge
transformation that is preserved by the background. Thus the
$U(1)$ symmetry is broken to $\Z_{2N}$.

\subsection{$U(1) \to \Z_{2N}$ in $G_2$ holonomy backgrounds?}

Can this argument be adapted to the $G_2$ holonomy case we are
studying? We will show that, classically at least, this is not
possible. The natural adaptation of the logic would be to have a
nonvanishing three-form field $C$ such that under a shift in the
massive isometry direction, $\psi
\to \psi + \eps$, one had $\delta C
\sim N \eps \w_3$. Where $\w_3$ would be the volume form of one of
the $S^3$s in the $G_2$ metric. The shift $\eps$ would then become
quantised using the same argument as before, but with instantonic
M theory membranes instead of D1-branes.

However, any $C$-field that we add classically must have
$G = d C = 0$ in order not to backreact and spoil the $G_2$
holonomy of the metric. The fact that $C$ must be closed means
that the periods of $C$ are invariant under homology. This will
not allow the desired periods of the form $\int C \sim N \psi$,
because two cycles at different values of $\psi$ are homologous
but would have different periods, contradicting our previous
statement.

The only way of introducing a $\psi$ dependence into $C$ without
having the periods proportional to $\psi$ is to integrate over
$\psi$ also. This would require a term like $C \sim \psi d\psi
\wedge \w_2$, for some closed two-form $\w_2$.
Such a term will also not work, because as $\psi \to \psi +
\eps$ then $\delta C = \eps d\psi \wedge \w_2 = d (\eps \psi \wedge \w_2
)$. Thus the change in $C$ is exact, and will not cause a change
in phase of the membrane partition function.

One should then investigate whether quantum mechanical effects may
alter the situation. Atiyah and Witten have shown
\cite{Atiyah:2001qf} that M theoretic corrections to $G_2$
holonomy vacua are under some control, and in particular that
nonvanishing four-form fluxes $\int G$ are induced. A na\"ive
application of the results in \cite{Atiyah:2001qf}, which use the
old $G_2$ metrics, does not look as promising as it might. Let us
sketch why.

Atiyah and Witten parameterise the space of asymptotically $G_2$
vacua in terms of three complex variables $\eta_1,
\eta_2, \eta_3$. The only property of these variables that we will
require is that
\be
\a_i \equiv \arg \eta_i = \int_{D_i} C + \mu(D_i) \pi\,,
\ee
where $D_i$ is a three-cycle of the geometry at infinity. The
second term, $\mu(D_i)$ is a topological correction due a membrane
fermion anomaly, it can be 0 or 1. There are three such cycles if
one thinks of the orbits of the cohomogeneity one $G_2$ metrics as
\be
\frac{S^3}{\Z_N} \times S^3 = \frac{SU(2)/\Z_N \times SU(2) \times SU(2)}{SU(2)} .
\ee
The cycles satisfy the constraint $N D_1 + D_2 + D_3 = 0$. A major
result of \cite{Atiyah:2001qf} is that the space of supersymmetric
vacua that are asymptotically $G_2$, including M theory
corrections, is given by the following equations
\bea\label{eq:atiyahwitten}
\eta_2 = \eta_1^{-N} (\eta_1 - 1)^N \,, \nonumber \\
\eta_3 = (1- \eta_1)^{-N} \,.
\eea
These equations parameterise a one complex dimensional space of
supersymmetric vacua that interpolates smoothly between three
classical regimes. These classical regimes are distinguished by
which of the three cycles at infinity, $D_i$, is filled in by the
full $G_2$ metric. In the $\dmet$ metrics, the cycle with the
$\Z_N$ quotient does not collapse in the interior, so it is
similar to the classical points in the space of vacua $P_2$ and
$P_3$ considered by \cite{Atiyah:2001qf}, in which $D_2$ and $D_3$
collapse respectively. Without loss of generality we take it to be
$P_3$. Note that \cite{Atiyah:2001qf} did not consider the $\dmet$
metrics, as they were working with the older $G_2$ holonomy
metrics. In this sense the present analysis is only preliminary,
and could change substantially if one redid the calculations of
Atiyah and Witten with the new metrics.

In the classical small curvature limit, that is, for the point
$P_3$, one has
\be
\int_{D_3} C = \int_{\R^4} G = 0.
\ee
One can also show \cite{Atiyah:2001qf} that $\mu(D_3)=0$ in this
case, so $\a_3 = 0$. Further, it turns out that as we tend towards
$P_3$, then $|\eta_1| \to 0$. The equations
(\ref{eq:atiyahwitten}) now imply that in the classical limit
\be
\a_2 = - N \a_1 + N \pi \,.
\ee
The periods of the $C$ field are also related because $N D_1 + D_2 +
D_3 = 0$ implies that
\be
\int_{D_2} C = - N \int_{D_1} C \,.
\ee
As argued above, the classical situation is not interesting
because the $C$ field cannot have the required $\psi$ dependence.
The quantum corrections to the $C$ field may be calculated from
(\ref{eq:atiyahwitten}). Take $\eta_1 = |\eta_1| e^{i \Theta}$,
with $|\eta_1| \ll 1$ so that we are near the classical limit.
Then the quantum corrected periods to leading order away from the
classical point $P_3$ are
\bea\label{eq:quantumcorrections}
\int_{D_1} C & = & \Theta \,, \nonumber \\
\int_{D_2} C & = & - N \Theta - N |\eta_1| \sin\Theta \,, \nonumber \\
\int_{D_3} C & = & N |\eta_1| \sin\Theta \,.
\eea
Whilst the appearance of factors of $N$ might appear promising,
these formulae do not allow the desired effect. Firstly,
$|\eta_1|$ and $\Theta$ are parameterising the space of vacua, so
they cannot depend on the coordinate $\psi$, they are constant for
a given background. Secondly, the equations
(\ref{eq:quantumcorrections}) tie the quantum corrected periods to
the classical periods. Thus the quantum corrections cannot change
under $\psi \to \psi + \eps$ without the classical periods also
changing. But we established above that the classical periods
could not change.

Nonetheless, to test this properly, one should extend the full
analysis to the newer $G_2$ metrics that we have been studying.
This seems like a worthwhile calculation to carry out in any case.

A remnant of the $U(1) \to \Z_{2N}$ breaking is seen in the
existence of $N$ vacua in the IR, where $\Z_{2N} \to \Z_2$. The
topological studies of confining strings
\cite{Acharya:2000gb,Acharya:2001hq} and domain walls
\cite{Acharya:2000gb,Acharya:2001dz} in the IR of the $G_2$
metrics support this picture.

An alternative argument for seeing $U(1) \to \Z_{2N}$ in $G_2$
backgrounds was suggested by \cite{Acharya:1998pm}. This involved
a four-form characteristic class $\lambda = p_1(G_2)/2$. It seems
unlikely that this argument can be applied here because
$H^4(G_2,\Z) = 0$, implying $\int_{\Sigma_4} \lambda = 0$, for any
four-cycle $\Sigma_4$.

Another way to address the problem is to reduce the $G_2$ solution
to a type IIA background. These have a nonvanishing Ramond-Ramond
one-form potential from the nontrivial fibration over the M theory
circle $\pa_{\gamma}$. One might have hoped to use a similar
argument to the MN background by considering the coupling of the
one-form potential to D0-branes. However, this will not work
because the IIA metric at infinity has vanishing first cohomology
$H^1(S^2\times S^3,\Z) = 0$. This implies that there are no gauge
transformations of the one-form potential that are not exact.
Therefore none of them will be discretised through quantum
mechanical effects. This should be contrasted with the MN situation in
which $H^2(S^2\times S^3,\Z) = \Z$ is the relevant cohomology group.

The conclusion of this subsection is that the $U(1) \to \Z_{2N}$
breaking appears to be more subtle in the $G_2$ backgrounds than
in, say, the MN solution. Although the effects of the breaking
seem to be visible indirectly in the IR, it seems the classical
background cannot cause the breaking directly in the UV. An
optimistic scenario is that the breaking may be visible after
considering M theory corrections to the background {\it \`{a} la}
Atiyah-Witten \cite{Atiyah:2001qf}. A pessimistic scenario is that
the effects of non-decoupling of gravity are worse for the
D6-branes than the non-decoupling of stringy modes for the
D5-branes, and therefore that the contamination of the pure gauge
theory in the UV results in an explicit rather than anomalous
breaking of the $U(1)$ symmetry.

\section{Massive isometries: masses from gravitational backgrounds}

{\it In this section we develop a theory of massive isometry in a
more general context. We show that another example of a massive
isometry is found in the Atiyah-Hitchin metric. We give a
covariant definition of the mass associated with a massive
isometry.}

\vspace{0.5cm}

\noindent We have argued that spontaneous symmetry breaking along a certain
angle in a $G_2$ holonomy metric is dual to the anomalous chiral
$U(1)$ R-symmetry of $\Ncal = 1$ super Yang-Mills theory. One
reason for selecting the particular angle is that fluctuations
about this, non-isometric, direction give a reduced Lagrangian
describing a massive vector field. Such an effect is known to be
generically dual to an anomalous chiral symmetry, although this is
the first time that the only background field is the metric, and
therefore the angle cannot be an isometry. We call a direction
about which one generates a mass term, and about which the
fluctuation is consistent to first order, a massive isometry.

We would like massive isometries to be fairly special directions.
Otherwise, they do not provide a sharp criterion for identifying
appropriate angles. To this end we now study fluctuations about a
generic non-isometric direction.

\subsection{Fluctuations about a generic direction}
\label{sec:genfluc}
Consider the metric
\be
\label{eq:metric}
ds^2 \equiv G_{u v} dx^u dx^v = h_{mn}(x,\psi) dx^m dx^n + \p^2(x)
\left( d\psi + K(x) + W(x) \right)^2 ,
\ee
where $K$ is part of the `background' metric, and $W$ is the
fluctuation. The use of indices is now that $m,n \ldots$ to run over
all coordinates except $\psi$.
One could also add a perturbation $\lambda(x)$ to $\psi$,
but this may be absorbed into $W(x)$ through the change of
coordinates/gauge transformation discussed above in
(\ref{eq:gauge}). Note the $\psi$-dependence in $h_{mn}$ which means
that this direction is not an isometry. We could have further
allowed $\phi$ and $K$ to depend on $\psi$, but this will not be
necessary for the class of metrics we are interested in.

The metric (\ref{eq:metric}) has vielbeins
\be\label{eq:generic}
e^\mh = e^\mh{}_n(x,\psi) dx^n \;\;\;\; e^\zh = \p \left(d\psi +
K + W\right) ,
\ee
and inverse vielbeins
\be
dx^m = e^m{}_\nh e^\nh \;\;\;\; d\psi = \p^{-1} e^\zh - K - W .
\ee

What is the action describing the fluctuation about a generic non-isometric
direction, as in (\ref{eq:metric})? We will now calculate this action.

In Appendix A we calculate the spin connections for this metric
(\ref{eq:metric}) and the Ricci scalar, in order to obtain the action for the
fluctuation by substituting into the Einstein-Hilbert action. The
result is
\be\label{eq:simpleaction}
R(W) = - \frac{1}{4} \p^2 F_{\sh \mh} F_{\sh \mh} + \left[
2 B_{\mh \sh} B_{\nh \sh} - 2 B_{\mh \nh} E + \left(E^2 - B_{\sh
\th} B_{\sh \th} \right) \d_{\mh \nh} \right] W_\mh W_\nh .
\ee
In this expression we have introduced quantities defined in
terms of the vielbeins (\ref{eq:generic}). These are
\bea\label{eq3}
E_{\mh \nh} & = & e^p{}_\mh \pa_\psi e^\nh{}_p , \nonumber \\
B_{\mh \nh} & = & E_{(\mh \nh)} , \nonumber \\
E & = & E_{\mh \mh} .
\eea
The field strength is $F = dW$.

Therefore, we find that mass terms are generic. There are no linear
terms because we are perturbing about a solution to Einstein's
equations, so the first order change to the action vanishes.
However, generating a mass was only
the first half of the definition of massive isometry. In order for
the perturbation to actually exist classically, it must be
consistent, that is, it must solve the full equations of motion.
We saw above in studying $G_2$ metrics, and will see in more generality below, that
this condition is indeed rather nontrivial.

\subsection{Application to the $G_2$ cases}

As a test of our formula, we may recover the action for the
$G_2$ metrics we discussed previously. To do this, we work with
the more specialised metric form
\be\label{eq:special}
ds^2 = dx^2_4 + dr^2 + e^\mh(r,y,\psi) e^\mh(r,y,\psi) + \phi(y,r)^2\left[d\psi
+ K(r,y)+W(x,r)\right]^2 \,,
\ee
where $W(x,r)$ will be the perturbation. We further require that
$K$ has no $dr$ component. Consistently with our previous index
notation, we let $a,b,\ldots$ run over the directions $t,x,y,z,r$
and let $m,n,\ldots$ run over the angular directions in the $G_2$
manifold, which are denoted $y$ in (\ref{eq:special}). The $G_2$
metrics have, for example,
\be
\phi(y,r)  =  u(r) \,, \quad K =
p(r) d\gamma + q(r) \cos\theta \, d\phi + s(r) \cos\alpha \,
d\beta \,,
\ee
where we just compare the metric (\ref{eq:special}) with the $G_2$
metric in the form (\ref{eq:g2metric}).

The various terms in the action may now be computed
\bea
E & = & 0 \,, \nonumber\\
B_{\mh \nh} B_{\mh \nh} &= & \frac{[b(r)^2-a(r)^2]^2}{4a(r)^2
b(r)^2} \,, \nonumber \\
\phi^2 & = & u^2 = c^2 + \frac{(g^2-1)^2 [2 a b f + c(a^2-b^2)]^2}{4 a^2 b^2} \,.
\eea
Substituting into a suitably specialised version of the expression
for the Ricci scalar (\ref{eq:simpleaction}) one obtains
\be
\left(\frac{(g^2-1)^2 [2abf + c(a^2-b^2)]^2}{64 a^2 b^2} + \frac{c^2}{4} \right) F_{a b}
  F^{a b} + \frac{[a^2-b^2]^2}{4a^2 b^2} W_a W^a ,
\ee
We also need the volume factor of the metric, which is
\be
(-G)^{1/2} = [ g_7(r) ]^{1/2} \sin\theta \sin\alpha .
\ee
Putting these together, and doing the integral over all the
angles, we reproduce the Lagrangian found before in
(\ref{eq:Waction}). Note that the action is Lorentz invariant. This
follows from the specialised ansatz (\ref{eq:special}).

\subsection{Application to the Atiyah-Hitchin metric}

The Atiyah-Hitchin metric arises in many contexts in string/M
theory \cite{Hanany:2000fw}. It was initially considered as the
moduli space space of two $SU(2)$ BPS monopoles in four dimensions
\cite{Atiyah:dv}. More relevant here will be the fact that when
extended with seven flat directions to be a supersymmetric
solution of M theory, the background is the strong coupling dual
of type IIA string theory on an $O6^-$ plane. It also describes
the Coloumb branch of a certain three-dimensional supersymmetric
gauge theory.

The metric may be written in the following Bianchi IX form,
\be\label{eq:atiyah}
ds^2_{11} = dx^2_7 + a(r)^2 b(r)^2 c(r)^2 dr^2 + a(r)^2 \s_1^2 + b(r)^2
\s_2^2 + c(r)^2 \s_3^2 .
\ee
The $\s_i$ are, as previously, left invariant one-forms of $SU(2)$
with Euler angles $(\theta,\phi,\psi)$. The radial functions satisfy the
following first order equations
\be
\dot{a} = \frac{a (b^2 + c^2 - a^2)}{2} - abc, \quad \text{+ cyclic.}
\ee
The ranges of the coordinates are $0\leq \theta \leq \pi$, $0 \leq
\phi < 2\pi$ and $0 \leq \psi < 2\pi$, so the symmetry of the manifold is
$SO(3)$ rather than $SU(2)$.

As $r\to\infty$ the metric acquires a further isometry, which is the
$U(1)$ generated by $\partial_{\psi}$. At finite $r$, this isometry is
broken by an exponentially small value of $a(r)^2 - b(r)^2$. The lack of
isometry in the interior of the metric has the important physical
consequence of nonconservation of charge of the individual monopoles
in two monopole scattering \cite{Gibbons:df}. Of course the total
charge is conserved. The corrections should be thought of as tree
level exchange of massive gauge bosons. These gauge bosons are the
perturbative degrees of freedom of the Higgsed field theory in which the
monopoles exist, they have a constant mass and should not be confused
with the massive gauge fields to be described shortly. When the
monopole scattering is embedded into string theory as intersecting
branes, the exponential corrections admit an elegant interpretation in
terms of instanton corrections \cite{Hanany:2000fw}.

The behaviour of the direction $\pa_\psi$ is thus very similar to what
we found for $\pa_\psi$ in the $G_2$ holonomy metrics. We will now
show that fluctuations about this direction may consistently be
described as a massive gauge field. Thus the Atiyah-Hitchin manifold
has a massive isometry.

As usual, we work in the $\lambda=0$ gauge and therefore the
fluctuation is just
\be
d\psi \to d\psi + W_a(x,r) dx^a .
\ee

Substituting into the eleven dimensional Eintein-Hilbert action we
find the following action for the fluctuation
\be\label{eq:ahaction}
S \propto - \int d^7x dr \left[ \frac{a^2 b^2 c^4}{4} F_{a b} F^{a
b} + \frac{(a^2-b^2)^2 c^2}{2} W_a W^a \right] .
\ee
The similarity with the $g=1$ metrics of $G_2$ holonomy is
remarkable. Note that in eight dimensions, indices are raised and lowered
with the metric (\ref{eq:atiyah}) with its nontrivial $g^{rr}$ coefficient.

We now need to check that this reduced action is consistent. This
is done similarly to before. We let
\be
P(r) = \frac{a^2 b^2 c^4}{4}, \qquad Q(r) = \frac{(a^2-b^2)^2 c^2}{2} .
\ee
The equations of motion following from the action (\ref{eq:ahaction})
are then
\bea
\pa_j F^j{}_r = \frac{Q}{2P} W_r \,, \nonumber \\
\pa_r (g^{rr} F_r{}^i) + \pa_j F^{ji} + \frac{\pa_r P}{P} g^{rr}
F_r{}^i = \frac{Q}{2P} W^i \,.
\eea
These imply the divergence equation
\be
\pa_i W^i + \pa_r (g^{rr} W_r) + \left[\frac{\pa_r P}{P} +
\frac{2P}{Q}\pa_r \left(\frac{Q}{2P} \right) \right] g^{rr} W_r = 0 .
\ee

We may now calculate the fluctuation of the eleven dimensional Ricci
tensor. Using the equations of motion we just derived and also the
first order equations for the radial functions, we find that
\be
\delta R_{\bar{u} \bar{v}} = 0 .
\ee
Therefore the reduced equations imply the full eleven dimensional
linearised equations of motion and the fluctuation is consistent.

It is interesting that again the first order radial equations are
necessary, showing that supersymmetry of the background is what allows
consistency of the fluctuation.

\subsection{Consistency in the generic case}

In this subsection we examine consistency of a generic massive
isometry. It follows from (\ref{eq:simpleaction}) that the generic
action for a massive isometry is
\be
S \propto - \int dx e \phi \left[ \frac{\phi^2}{4} F_{mn} F^{m n}
- \frac{n-2}{n} \left(E^2 - B_{\sh \th} B_{\sh \th} \right) W_m
W^m \right] ,
\ee
where $e = \det e^\mh{}_k$. Here $n$ denotes the number of $m,n,\ldots$
indices. Note that we are now working in curved
indices for the gauge fields. We have integrated out the $\psi$
angle, so it is important that the Lagrangian does not depend on
this direction, except possibly as an overall factor. The equation
of motion following from this action is then
\be
\nabla_n F^{n m} + \frac{2(n-2)}{n} \frac{E^2-B_{\sh \th} B_{\sh
\th}}{\phi^2}A^m + \left[ 3 \frac{\pa_n \phi}{\phi} + e^s{}_\th \pa_n e^\th{}_s
\right] F^{n m} = 0 .
\ee
The equation is perhaps clearer in the case of the restricted
ansatz of equation (\ref{eq:special}). In this case the equations
of motion are
\be\label{eq:fluctuationequations}
\pa_a F^{a b} + 2 \frac{E^2-B_{\mh \nh} B_{\mh
\nh}}{\phi^2}A^b + \left[ 3 \frac{\pa_r \phi}{\phi} +
e^s{}_\th \pa_r e^\th{}_s \right] F^{r b} = 0 ,
\ee
where $a,b$ run over the five directions $t,x,y,z,r$.

Working in case of (\ref{eq:special}), we now calculate the
linear terms in the fluctuation for the Ricci tensor. These must all
vanish for the perturbation to be consistent. Let us look at some of
the simpler components
\bea
\d R_{\bar{\psi}\bar{\psi}} & = & \frac{\pa_r\p}{\phi} E W_{\bar{r}} \,,
\nonumber \\
\d R_{\mh\bar{\psi}} &= & \frac{1}{2} \phi W_{\rh} \left( E_{\sh\mh} H_{\sh
  \rh}+ E H_{\rh\mh} + \pa_\psi H_{\rh\mh} \right) \,, \nonumber \\
\d R_{\bar{r}\bar{r}} & = & - \left(2 B_{\mh \nh} A_{\rh \mh \nh} + \pa_\psi A_{\rh \sh \sh} \right)
W_rh + \pa_r (E W_\rh) \,, \nonumber \\
\d R_{\ih\bar{r}} & = & \frac{E}{2} \left( \pa_i W_\rh + \pa_r W_{\ih} \right)
- W_{\ih} B_{\mh\nh} A_{\rh\mh\nh} \,, \nonumber \\
\d R_{\ih\bar{\psi}} & = &  - \frac{\phi}{2} \left[\pa_{\bar{c}}
  F_{\bar{c} \ih} + F_{\rh \ih} \left(3 \frac{\pa_r
    \phi}{\phi} + e^s{}_{\mh} \pa_r e^\mh{}_s \right) + \frac{2 (E^2 -
    B_{\nh\mh}B_{\nh\mh})}{\p^2}
W_{\ih} \right] + \frac{W_{\ih}}{\phi} \left[\pa_\psi E + E^2
  \right]\,, \nonumber \\
& = & \frac{W_{\ih}}{\phi} \left[\pa_\psi E + E^2 \right] \,.
\eea
Note that in the last of these equations we have used the equations
of motion for the fluctuation (\ref{eq:fluctuationequations}).
In these expressions we have used the objects
\bea
A_{\mh\nh\sh} & = & -e^p{}_{[\mh} e^q{}_{\nh]}
\pa_p e^\sh{}_q - e^p{}_{[\mh} e^q{}_{\sh]} \pa_p e^\nh{}_q +
e^p{}_{[\nh} e^q{}_{\sh]} \pa_p
e^\mh{}_q , \nonumber \\
H & = & d K \,,
\eea
where $K$ is defined in equation (\ref{eq:metric}).
We see that in order for these fluctuations to vanish, we need the following independent
consistency conditions:
\be\label{eq:con1}
E=0\,,
\ee
\be\label{eq:condition}
E_{\sh \mh} H_{\sh \rh} + \pa_\psi H_{\rh \mh} =0\,.
\ee
\be\label{eq:con2}
A_{\rh\sh\nh}B_{\sh\nh} =0 \,,
\ee
and
\be\label{eq:con3}
\6_\psi A_{\rh\sh\sh} =0\,.
\ee

There are, of course, more consistency conditions which follow from requiring the vanishing of other
$\d R_{\bar{\mu}\bar{\nu}}$ terms.  These are in general fairly long
and unilluminating. One further simple condition that may be found comes from
\be
\frac{W_{\ih}}{W_\rh} \d R_{\rh \nh} - \d R_{\ih \nh} = \frac{\phi^2}{2}
H_{\rh \nh} F_{\rh \ih} \, ,
\ee
which then implies
\be\label{eq:con5}
H_{\rh \nh} = 0 .
\ee
Note that this is sharper than, and implies, equation (\ref{eq:condition}) above.

We may apply these formulae to the $G_2$ metrics. We find that the
first four conditions, (\ref{eq:con1}) to (\ref{eq:con3}), are
satisfied by all the $G_2$ holonomy metrics. The fifth condition
(\ref{eq:con5}) is satisfied when the metric function $g(r)=1$,
just as we found previously in section \ref{sec:consistency}. In
fact, calculating the right hand side of (\ref{eq:con5})
explicitly, one finds precise agreement with
(\ref{eq:inconsistent}). This gives a pleasant check on the
calculations of this and that section.

It is hopefully apparent from the simpler conditions for
consistency we have derived here, and the more complicated
conditions we have not discussed, that the requirement that a
massive vector fluctuation be consistent, and hence a massive
isometry, is nontrivial. The necessary use of the first order
supersymmetry equations for the background, see section
\ref{sec:testI}, further supports this statement. Asymptotic
consistency, that is, consistency in the $g(r)=1$ $G_2$ metrics,
is also significantly nontrivial. In particular, there does not
seem to be another obvious non-isometric angle in the $g(r)=1$
metrics that would satisfy the condition (\ref{eq:con5}).

\subsection{Towards a covariant description of massive isometry}

In previous subsections, we used a particular choice of coordinates in
which we perturbed the background solution. The definition we gave
above of massive isometry depended on first finding coordinates such
that the metric takes on a specific form.

An interesting question is whether a fully covariant description of
massive isometry is possible. That is, given a non-isometric direction
with tangent vector $k$, if one takes coordinates adapted to this vector
field
\be
k = \frac{\pa}{\pa \psi} ,
\ee
is there a collection of covariant conditions that $k$ must satisfy in
order for the perturbation $d\psi \to d\psi + A$ to produce a massive
isometry?

Such a description would be analogous to the Killing vector
equation for an isometry. If we have a vector field $k$ satisfying
$\nabla_u k_v + \nabla_v k_u = 0$, then it is well known that if we
take coordinates adapted to this vector field and perturb, then the
fluctuation is described by massless electrodynamics.

We will not attempt to find a fully covariant description of
massive isometry here. However, we will give a covariant
description of the mass of the vector field associated with a
massive isometry.

Let $k$ be a massive isometry, consider
\be
\nabla_\uh k_\vh +\nabla_\vh k_\uh = e^s{}_\uh \pa_s k_\vh + e^s{}_\vh
\pa_s k_\uh + \left[w_{\uh\vh\sh} + w_{\vh\uh\sh} \right] k_\sh .
\ee
If we work in the adapated coordinates, assumed to be of the form of
(\ref{eq:metric}), then we
have $k_\zh = \phi$ and $k_\mh = 0$. Then, using the spin connections
of (\ref{eq:spinA}) in the Appendix we find that
\be
\nabla_\zh k_\zh = 0 \,, \qquad \nabla_{(\zh} k_{\mh )} = 0\,, \qquad
\nabla_{(\mh} k_{\nh )} = B_{\mh\nh} ,
\ee
where $B_{\mh\nh}$ is defined in equation (\ref{eq3}).
Note that this result neatly includes the case when $k$ is an isometry
and $B_{\mh\nh}$ is zero. The expression is not tensorial however.

Another expression that may be derived similarly is
\be
\nabla_\uh k_\uh = E =  E_{\mh\mh}.
\ee
Combining these results and refering to equation (\ref{eq:simpleaction}) we see
that the mass of the vector field will be proportional to
\be
\text{mass}^2 \propto \frac{B_{\mh\nh}B_{\mh\nh} - E^2}{\phi^2} = - \frac{k_\uh
\nabla_\vh \nabla_\vh k_\uh + k_\uh
\nabla_\vh \nabla_\uh k_\vh + 2 \left( \nabla_\uh \right) \left( k_\uh
\nabla_\vh k_\vh\right)}{2 k_\uh
k_\uh} .
\ee
This result takes on a particularly nice form in the case when
$E=0$, and when the background is Ricci flat. Both these conditions
are satisfied for the $G_2$ metrics. As we have seen, the condition $E=0$ is
in fact necessary for a massive isometry to be consistent.
Ricci flatness allows us to use $\nabla_\uh
\nabla_\vh k_\uh = \nabla_\vh \nabla_\uh k_\uh$. If these conditions
hold then one obtains the following simple expression for the mass
\be
\text{mass}^2 \propto \frac{B_{\mh\nh}B_{\mh\nh}}{\phi^2} = - \frac{k \cdot
\nabla^2 k}{2 k^2} .
\ee

\section{Conclusions and future directions}

The main conclusion of our work is that $G_2$ holonomy metrics exhibit
spontaneous symmetry breaking of a vector fluctuation about a certain
non-isometric angle. This is an important phenomenon within the context
of gauge-gravity dualities, because on general grounds anomalous global
symmetries in field theory are expected to be dual to spontaneously
broken gauge symmetries in gravity. In the $G_2$ holonomy case, we
have argued that the vector fluctuation we considered is dual to the
anomalous chiral current of ${\mathcal{N}}=1$ super Yang Mills
theory. This suggests that $G_2$ holonomy duality may have a
dictionary with similar structure to other better studied
dualities, even though the background is not asymptotically Anti-de
Sitter. The further elucidation of this dictionary is an interesting
question. Can the identification of a vector fluctuation as dual to the
chiral current be pushed further to allow the computation of gauge
theory two point functions?

We have argued that the appropriate $G_2$ holonomy metrics to use
are the fairly recently constructed $\dmet$ metrics. The
availability of concrete metrics, up to radial functions specified
by ODEs, was crucial for our work. These metrics should allow
further matchings between M theory on the $G_2$ manifolds and
${\mathcal{N}}=1$ gauge theory. Perhaps one can adapt to the $G_2$
duality the recent success in calculating the ${\mathcal{N}}=1$
beta function from the Maldacena-N\'u\~{n}ez background
\cite{DiVecchia:2002ks,Bertolini:2002yr}? The concrete metrics
should also be useful in the IR of the field theory. So far the
topological charges of confining strings and domain walls have
been matched. The existence of a metric should allow further
elucidation of the quantitative dynamics of the strongly coupled
regime.

We found that the fluctuation we studied had a five dimensional
interpretation that was consistent with previous results. However,
it would be useful to understand the five dimensional perspective
better by embedding the fluctuation into a full five dimensional
supergravity, on a background that lifted to the $G_2$ holonomy
metrics. It seems plausible that such a supergravity could be
constructed by reducing the eleven dimensional theory on $SU(2)$
to get an eight dimensional supergravity, as has already been done
to construct $G_2$ holonomy metrics
\cite{Edelstein:2001pu,Hernandez:2002fb,Edelstein:2002zy}, and
then doing a further reduction on another $SU(2)$ to get to five
dimensions.

A question that needs further clarification is the breaking of
$U(1) \to \Z_{2N}$. It seems that this is not visible from the
classical supergravity solution in the UV. We made some
preliminary comments above about whether quantum corrections to
the classical $G_2$ background, based on the work of Atiyah and
Witten \cite{Atiyah:2001qf}, would induce a $C$ field with the
correct structure to cause the $U(1) \to \Z_{2N}$ breaking. It
appears that a systematic generalisation of their work to the new
$G_2$ holonomy metrics is required to settle the issue.

Another set of questions thrown up by this work concerns the
notion of massive isometry. This concept is natural when
considering duality in purely gravitational backgrounds with a
broken symmetry. Is a covariant description of massive isometry
possible? Can one find general conditions for consistency of
massive isometries? Does the notion have other applications,
perhaps phenomenological? A final issue that would be nice to
clarify is the relation between consistency of the fluctuation and
supersymmetry of the background that we found above.

\vspace{1.5cm}
\centerline{\bf Acknowledgements}

We are very grateful to Carlos Nu\~{n}ez for stimulating our
interest in dual descriptions of chiral symmetry and for many
helpful suggestions throughout this work. We have further had
useful conversations with Oliver DeWolfe, Dan Freedman, Gary
Gibbons, Robert Helling, Juan Maldacena, Giusseppe Policastro,
Fernando Quevedo and David Tong.

S.A.H and R.P. would like to thank the Center for Theoretical Physics, MIT,
for hospitality while part of this work was carried out.

The work of the authors is financially supported by the Sims
scholarship (S.A.H.), Trinity College Cambridge (R.P.) and the
D.O.E. cooperative research agreement \#DF-FC02-94ER40818 (U.G.).

\appendix

\section{Calculating the action for a generic fluctuation}

This appendix evaluates the Ricci scalar of the metric
\be
\label{eq:metricA}
ds^2 \equiv G_{u v} dx^u dx^v = h_{mn}(x,\psi) dx^m dx^n + \p^2(x)
\left( d\psi + A(x) \right)^2 ,
\ee
that was of interest in section \ref{sec:genfluc}. Here we have
set $A = K + W$, where $K$ is part of the `background' metric, and
$W$ is the fluctuation. The indices $m,n \ldots$ run over all
coordinates except $\psi$.

The metric (\ref{eq:metricA}) has vielbeins
\be\label{eq:genericA}
e^\mh = e^\mh{}_n(x,\psi) dx^n \;\;\;\; e^\zh = \p \left(d\psi +
A\right) ,
\ee
and inverse vielbeins
\be
dx^m = e^m{}_\nh e^\nh \;\;\;\; d\psi = \p^{-1} e^\zh - A .
\ee

One may now calculate the spin connections. We find
\bea\label{eq:spinA}
\w_{\mh \nh} & = & \left[ A_{\mh \nh \sh} + B_{\sh \nh} A_{\mh} - B_{\sh \mh}
A_{\nh} - C_{\mh \nh} A_{\sh}\right] e^\sh + \left[ C_{\mh \nh} -
\frac{1}{2} \p^2 F_{\mh \nh} \right]\p^{-1} e^{\zh} , \nonumber \\
\w_{\zh \mh} & = & \left[\frac{1}{2} \p^2 F_{\mh \nh} - B_{\mh \nh}
\right] \p^{-1} e^\nh + D_\mh \p^{-1} e^\zh .
\eea
Where $F = dA$ and we have introduced the following quantities,
\bea\label{eq3A}
A_{\mh\nh\sh} & = & -e^p{}_{[\mh} e^q{}_{\nh]}
\pa_p e^\sh{}_q - e^p{}_{[\mh} e^q{}_{\sh]} \pa_p e^\nh{}_q +
e^p{}_{[\nh} e^q{}_{\sh]} \pa_p
e^\mh{}_q , \nonumber \\
E_{\mh \nh} & = & e^p{}_\mh \pa_\psi e^\nh{}_p \,, \nonumber \\
B_{\mh \nh} & = & E_{(\mh \nh)} , \qquad \qquad
C_{\mh \nh} = E_{[\mh \nh]} , \nonumber \\
D_\mh & = & e^p{}_\mh \pa_p \p .
\eea
These quantities may be used to calculate a reduced action by
substituting into the Einstein-Hilbert action. After neglecting a
total derivative, the Einstein-Hilbert action may be written in
terms of the spin connection as follows
\be
\label{spinactionA}
S \propto \int dx d\psi G^{1/2} \left[ \w^\uh{}_{\vh\sh}
\w^{\vh\sh}{}_\uh + \w^\uh{}_{\uh \vh} \w^\sh{}_\sh{}^\vh \right]
.
\ee
Where
\be
\w_{\vh\sh} = \w^\uh{}_{\vh\sh} e^{\uh} .
\ee
For our metric ansatz (\ref{eq:metricA}) we have $G^{1/2} = h^{1/2}
\p$ and the spin connection terms give
\be
\label{eq2A}
\w_{\sh \mh \nh} \w_{\mh \nh \sh} + \w_{\mh \mh \sh} \w_{\nh \nh \sh} +
2 \w_{\mh \zh \sh} \w_{\zh \sh \mh} + \w_{\mh \sh \zh} \w_{\sh \zh
\mh} + 2\w_{\mh \mh \sh} \w_{\zh \zh \sh} + \w_{\mh \mh \zh} \w_{\sh \sh
\zh},
\ee
where we have lowered all the flat indices using the flat metric.
This expression is found to be
\bea
\label{eq1A}
\p^{-2} \left[ E^2 - B_{\mh \sh} B_{\mh \sh} \right] + 2 \p^{-1}
D_{\sh} A_{\mh \sh \mh} + A_{\mh \sh \mh} A_{\nh \sh \nh} + A_{\mh
\nh \sh } A_{\nh \sh \mh} \nonumber \\
-\frac{1}{4} \p^2 F_{\sh \mh} F_{\sh \mh}
+ F_{\sh \mh} C_{\sh \mh} \nonumber \\
+ \left[2 \left(\p^{-1} D_\sh +  A_{\mh \sh \mh} \right)
\left( E_{\sh \nh} - E \d_{\sh \nh} \right) + 2 A_{\mh \nh \sh} E_{\mh
\sh}\right] A_{\nh} \nonumber \\
+ \left[ 2 B_{\mh \sh} B_{\nh \sh} - 2 B_{\mh \nh} E + \left(E^2 -
B_{\sh \th} B_{\sh \th} \right) \d_{\mh \nh} \right] A_\mh A_\nh.
\eea
We have introduced $E = E_{\mh \mh}$. Note that the expression
contains terms with no $A$ dependence, with linear and with
quadratic $A$ dependence, and with linear and quadratic $F$
dependence. Recall that $A = K + W$, where $K$ is part of the
background and $W$ is the fluctuation. Because we are perturbing
about a solution, we know that the linear terms in a perturbation of
the action must vanish.
Further, the zeroth-order terms may be collected into the
background Ricci scalar. Thus the result is in fact the simple
expression
\be\label{eq:simpleactionA}
R(w) = R(W=0) - \frac{1}{4} \p^2 F_{\sh \mh} F_{\sh \mh} + \left[
2 B_{\mh \sh} B_{\nh \sh} - 2 B_{\mh \nh} E + \left(E^2 - B_{\sh
\th} B_{\sh \th} \right) \d_{\mh \nh} \right] W_\mh W_\nh ,
\ee
where now we use $F = dW$. This is the result we quoted in (\ref{eq:simpleaction}).

\end{document}